TITLE

In-situ determination of the mechanical properties of gliding or non-motile bacteria by Atomic Force Microscopy under physiological conditions without immobilization.

AUTHORS

Samia Dhahri[1], Michel Ramonda[2] and Christian Marlière[1, 3]*

AFFILIATIONS

1. Géosciences Montpellier, University Montpellier 2, UMR CNRS 5243, Montpellier, France
2. Centrale de Technologie en Micro et nanoélectronique, Laboratoire de Microscopie en Champ Proche, University Montpellier 2, Montpellier, France
3. Institut des Sciences Moléculaires d'Orsay, ISMO, University Paris-Sud, UMR CNRS 8214, Orsay, France.

* CORRESPONDING AUTHOR

**Dr. Christian Marlière**
Institut des Sciences Moléculaires d'Orsay (ISMO), UMR CNRS 8214,
Bâtiment 350, Université Paris-Sud, 91405 Orsay Cedex, France,
Phone: (+33) 670 532 190 / email: christian.marliere@u-psud.fr

ABSTRACT

   We present a study about AFM imaging of living, moving or self-immobilized, bacteria in their genuine physiological liquid medium. No external immobilization protocol, neither chemical nor mechanical, was needed. For the first time, the native gliding movements of Gram-negative *Nostoc* cyanobacteria upon the surface, at speeds up to 900µm/h, were studied by AFM. This was possible thanks to an improved combination of a gentle sample preparation process and an AFM procedure based on fast and complete force-distance curves made at every pixel, drastically reducing lateral forces. No limitation in spatial resolution or imaging rate was detected.

   Gram-positive and non-motile *Rhodococcus wratislaviensis* bacteria were studied as well. From the approach curves, Young modulus and turgor pressure were measured for both strains at different gliding speeds and are ranging from 20±3 to 105±5MPa and 40±5 to 310±30kPa depending on the bacterium and the gliding speed. For *Nostoc*, spatially limited zones with higher values of stiffness were observed. The related spatial period is much higher than the mean length of *Nostoc* nodules. This was explained by an inhomogeneous mechanical activation of nodules in the cyanobacterium. We also observed the presence of a soft extra cellular matrix (ECM) around the *Nostoc* bacterium. Both strains left a track of polymeric slime with variable thicknesses. For *Rhodococcus,* it is equal to few hundreds of nanometers, likely to promote its adhesion to the sample. While gliding, the *Nostoc* secretes a slime layer the thickness of which is in the nanometer range and increases with the gliding speed. This result reinforces the hypothesis of a propulsion mechanism based, for *Nostoc* cyanobacteria, on ejection of slime.

These results open a large window on new studies of both dynamical phenomena of practical and fundamental interests such as the formation of biofilms and dynamic properties of bacteria in real physiological conditions.

INTRODUCTION

Atomic force microscopy (AFM) is a very powerful tool to get precious information about the nanoscale surface architecture of cells, the localization and interactions of their individual constituents with various internal and external agents. For instance, purified bacterial membranes have been imaged with very high resolution revealing the evolution of the organization of photosynthetic membranes in response to light [1]. Changes of cell surface structure were observed in physiological conditions during the germination of *Aspergillus fumigatus conidia* [2] at a resolution of a few nanometers. The nanomechanical properties of live cells have been mapped quantitatively using high rate dynamic AFM with a resolution in sub-10 nm range [3]. Recently, very high-speed atomic force microscopy with nanometer resolution revealed the action of an antimicrobial peptide on individual *Escherichia coli* cells [4].

One important common aspect of all these reported results is that the employed methods required to *immobilize* the cells firmly enough to enable them not to be swept out during the AFM scanning. The claimed aim was to withstand the lateral friction forces exerted by the tip during scanning without denaturing the cell interface and surface. As mentioned in many recent papers and reviews [5–7] that immobilization step has been considered as mandatory. In order to increase the adhesion of microorganisms, such as bacteria, to surfaces for AFM imaging in liquid, many protocols have been proposed. One of the first used methods required the drying of the sample [8], therefore limiting drastically cell viability [9]. Furthermore re-immersion in buffer medium was a challenging step. Many methods based on the use of various chemical ligands between the micro-organisms and the sample have been proposed such as polyphenolic proteins extracted from the marine mussel [10]. Another very common technique consists of pretreating the support with polycations, such as poly-L-lysine [11] or derivatives. However, the active groups used in covalent binding and the reagents used for cross-linking are known to affect cell viability by cross-linking the proteins on the cell surface [12] or by inducing cell porosity [13]. Mechanical entrapment in aluminum oxide filters [14] or in porous membranes such as isopore polycarbonate membranes [15] were developed to avoid slow poisoning by chemical immobilizing and proved to be effective. Such entrapment methods are mainly suitable for imaging and force measurements on spherical cells. However this technique may impede the monitoring of active processes such as cell division and bacteria may be in a state of mechanical stress far away from their standard living conditions. Furthermore, parameters such as pore size compatibility, depth of pores and confinement effects are difficult to master. Other physical methods based on trapping in soft gels as agar, gelatin layers [16] have been considered but these extracellular structures may lead to artifacts in AFM images through, for example, contamination of the apex of the tip. Recently, a new method has been proposed. It consists in assembling the living cells on specific areas within the patterns of micro-structured, functionalized poly-dimethylsiloxane stamps using convective/capillary deposition [17].

Because of immobilization step, using entrapment in membrane's pores or gluing techniques by means of chemicals, the cells are likely far away from their natural physiological conditions, which is a limiting factor for the biological relevance of such obtained results. Furthermore, by immobilization, one important aspect of many bacteria is overshadowed: Depending on the strain, bacteria may move inside the buffer medium by swimming or along limiting surfaces by gliding. In this last case, bacteria usually form the so-called biofilms [18]: structured, multicellular communities embedded in a matrix of extracellular polymeric substances (EPS) on solid substrates. Such biofilms are of very practical importance as they concern ubiquitous issues as in geophysical field [19] and is prevalent in natural, industrial and hospital settings [20], etc. An important parameter in the formation of biofilms is the mobility of bacteria. Many Gram-negative and Gram-positive bacteria have been shown to possess on their outer surfaces a variety of organelles known as pili, flagella or injectisomes useful for their motility [21–24]. However a large part of bacteria exhibits the ability to be motile along a surface without the aid of these organelles. *Nostocales* belong to this group and contains most of the species of cyanobacteria capable of gliding [25,26]. Gliding is a form of cell movement that does not rely on any obvious external organ or change in cell shape and occurs only in the presence of a solid sample [27,28]. The gliding mechanism itself has been mainly studied using conventional optical microscopy following the evolution of a pre-deposited seedling root on an agar gel. However, due to its gelatinous cell walls, bacteria adhere strongly to the agar gel [28]. Optical observations require the follow-up of the gliding movement during very long times, typically one day or more. The first moments of gliding and consecutively its mechanism are thus unreachable [29]. Different hypotheses about the origin of the gliding in filamentous cyanobacteria have been proposed: it could be powered by a ''slime jet'' mechanism, in which the cells extrude a gel through pores surrounding each cell septum providing a propulsion force [30–32]. An alternative hypothesis is that the cells use contractive elements that produce undulations running over the surface inside the slime tube like an earthworm [33]. A recent study, performed on immobilized bacteria (in dental wax) and based on the use of static AFM data [34] and scanning electron microscopy, revealed that arrays of parallel fibrils may be involved in the gliding mechanism. Also recent observations were made using new ellipsometric optical microscopy technics on *Myxococcus xanthus* [35]. The authors observed that slime is deposited at constant rate underneath the cell body slime. They suggested that this polymeric layer promotes the adhesion of the bacterium to the substrate without an effective participation to the motility machinery.

Here we present a study on AFM imaging of living bacteria in their standard (physiological) liquid environment without any external immobilization step. This was feasible thanks to an original method relying on an improved combination of a gentle sample preparation process and an AFM procedure based on fast and complete force-distance curves made at every pixel, drastically reducing lateral forces [36]. Two examples of very different strains are presented. First a non-motile, Gram-positive species (*Rhodococcus wratislaviensis,* IFP 2016, IFP Energies Nouvelles, Rueil-Malmaison, France) [37]. Highly spatially resolved AFM images of this species are obtained. Secondly, a filamentous, Gram-negative *Nostoc* cyanobacterium (*Anabaenopsis circularis,* PCC 6720, Institut Pasteur, Paris, France) well known for its gliding movement on surfaces of glass slides was studied. With this last bacterial strain, we were able to visualize for the first time the bacterium *during* its gliding movement on a glass slide by using AFM at a high acquisition rate (two images per minute and above) without limitation in spatial resolution. Different gliding speeds, from few tens to hundreds of nanometers per second, measured by cross-compared optical microscopy and AFM data, were studied. AFM height and mechanical stiffness data were simultaneously acquired. From these, mechanical parameters, inner turgor pressure and Young modulus, were derived. These results are a direct proof of the low impact of these breakthroughs AFM observations on the native behavior of the bacteria as its living character is verified thanks to its movement.

RESULTS

We successively studied two categories of bacteria: first *Rhodococcus wratislaviensis* [37], a bacterium currently used for degradation of mixtures of hydrocarbons, known to be non-motile, and, second, *Nostoc* bacteria able to glide on solid samples without the help of flagella or pili.

We first report results on *R. wratislaviensis.* Height images were taken in liquid phase (MM medium) by using QI mode [36] (see Material and Methods section) and are plotted in figure 1 in 2D (figure 1.a) and 3D (figure 1.b) using following parameters: 256 by 256 pixels, scan width 5 microns, acquisition time for one image around 10 minutes. The maximum applied force during the approach of the AFM tip was 6nN. The mean indentation depth in bacterium is around 40 nanometers. In the 3D image very fine topographic details on bacteria membranes such as small protuberances are better evidenced. It should be noted that the AFM image quality in liquid, moreover for non-immobilized bacteria, is unsurpassed [38,39]. The mean height of such bacteria is around 1 micron for a typical width of 1.2 microns. Two height profiles are plotted (figures 1.c and 1.d). In these profiles, a swelling (lateral expansion 450nm) is clearly visible at the right side of the main part of the bacteria and corresponds to a lower thickness (around 150nm) than on the central part of the bacterium. The only presence of these swellings at the right side of the bacteria is very unlikely due to an AFM scanning artifact. Indeed a smoothing of the height profile at the edge of the bacterium, due to a spurious convolution effect related to the finite value of semi-top angle of the AFM tip, can be very likely ruled out: such effect would have been preferentially located at lower parts of the bacteria as the cantilever, perpendicular to the horizontal side of the AFM image, is scanning from the top to the bottom. The approach curves giving the variation of the applied force versus the scanner elongation for the bacterium, the right swelling and the glass slide are plotted in figure 2. They revealed typical features for three different components with various stiffnesses: the softest one has a very rounded profile for low force domain and can be related to the bacterium's right swelling; the stiffest one with a linear behavior concerns the glass substrate; and the intermediate stiffness is characteristics of the bacterium.

The corresponding stiffness image, simultaneously acquired with topographic data, is shown in figure 1.e. The two stiffness profiles along the two scan lines as defined in height data are plotted in figures 1.f and 1.g. It must be pointed out that these values of stiffness are effective ones as they result from the association of two linear springs in series [40], one related the AFM's cantilever and the other to the bacterium envelope. The "glass slide" level corresponds to a stiffness of 0.35±0.01N/m, which is the typical stiffness value of the working cantilever. This is very near from the value we obtained by standard determination based on thermal noise [41] : $k$ = 0.36±0.01N/m. This value is the maximum stiffness we can measure with our AFM system as it works like a mechanical low-pass filter limited by the finite elasticity of the cantilever. This upper level is clearly evidenced by the light gray level in stiffness image. The effective stiffness of the bacterium has an approximate value of 0.23±0.05N/m. The swellings at the right side of bacteria, as defined above, are characterized by an effective stiffness of around 0.11±0.03N/m. These three main levels of stiffness are more visible in the histogram plot of the stiffness values (figure 3). Peak number 1 (P1) corresponds to the glass sample (0.35±0.01N/m), P2 to the bacterium (0.23±0.05N/m), and P3 to the swelling at the right of bacteria (0.12±0.02N/m). Please note that these effective stiffness values have been determined by using automatic peak detections (lorentzian shape assumed) through *OriginPro8.5* software (from *OriginLab*). That smaller value for effective stiffness (P3) is related to the presence of a polymeric layer, we will define it as *slime,* mainly located at the right side of the main part of the bacteria as observed in height images (figures 1.a, 1.b). The small protuberances on the membranes of *R. wratislaviensis* as seen in height data (figure 1.b) correspond to dark grey spots in stiffness images, zones with lower stiffness. We also investigated *R. wratislaviensis* with a higher lateral resolution as it is shown in figure 4 where new data (256x256 pixels; lateral scan size of 1 micron) were taken. Minute details on the membrane of the bacterium and the presence of a slime layer nearby the bacterium as well are clearly evidenced. In the histogram data (Fig. 4.d) of the stiffness image (figure 4.c) three peaks are visible and are positioned at stiffness values very similar to those previously determined: 0.35±0.02N/m (P3, the glass slide contribution : up/right corner of the image), 0.24±0.05N/m (P2, bacterium's membrane) and 0.10±0.02N/m (P1, slime). These results demonstrate that topographic and mechanical AFM data of high spatial resolution can be acquired by using our experimental protocol and set-up without the introduction of any external immobilizing step. Only the natural adherence of the bacterium on the sample through the slime layer contributes to the feasibility of the acquisition of such AFM data. No parasite sweeping movements of bacterium during the AFM scan were detected. These images are very stable in time as several acquisitions were done during at least two hours without any noticeable changes in conformation.

Thereafter cyanobacteria were studied. By means of optical microscopy we first observed the gliding movement of these filamentous *Nostocs* along the surface of the solid sample. Figure 5 is a typical example of such motility: the arrows indicate two successive positions of one of the nodule over time. This motility was optically observed regardless of the distance between the AFM cantilever and the substrate: from few microns (figure 5) to zero (AFM tip in contact). A typical curve showing the variation of the position of one of the *Nostoc* nodules (along the horizontal axis, *X*) is plotted in figure 6.a (tip/substrate distance = 500µm). As often noticed [27], forward and backward movements are observed. Same kind of movement for another *Nostoc* is optically observed when the AFM tip scanned the sample: the *Nostoc* displacement along *X* axis (figure 6.b, solid line) and its related *X* speed (figure 6.b, left triangles) are reported. The *Y* position of the AFM cantilever during its scanning movement is plotted in lower parts of figures 6.b and 6.c. The starting time of each AFM image is marked by a short vertical segment as indicated in figures 6.b and 6.c.

We will successively describe three typical sets of AFM data of the *Nostoc* cyanobacteria. The maximum applied force during the approach of the AFM tip was 10nN.

The first one is that mentioned above when describing surface displacements as measured by optical microscopy (figure 6). In figure 7 five successive AFM images (64X64 pixels; 40.3μmx7.6μm) are shown. In figure 7.6, an optical image (such as figure 5) was numerically modified to approximately get the same magnification as for AFM data. Figures 7.5 (AFM data) and 7.6 (optical data taken fifty microns far away from the investigated AFM scanning zone) remarkably revealed the same type of features: nodules with an average length of around 4 microns are evidenced with two small vertical bumps in front and back positions. The mean height and width of *Nostoc* are estimated to 2.2±0.2μm and 1.8±0.2μm respectively (figure 8.e). The *Nostoc* is clearly moving from left to right of the scanned region between images *1* and *2* (figure 7) before gliding in the opposite direction between images *2* to *5*. In figure 7.3 a slight distortion in the structure of the nodules is visible: there is indeed a shift of the left extremity of each nodule towards the left side of the image between its bottom and top (i.e. along the vertical scan direction). From these AFM data, the axial deformation across the width of the bacterium was measured and its value is equal to 1.5μm. As seen in figure 6.c, the bacterium is accelerating –as determined by optical data- during the acquisition of AFM image number 3 (maximum absolute speed around 500nm/s). As the mean width of the bacterium is around 2 microns (see below), the related AFM vertical scanning time of bacterium along its width is estimated to about 9 seconds. The resulting deformation is thus evaluated to 2±0.5 microns, very similar to the directly measured value (1.5±0.1μm). One important consequence of this consistency between AFM and optical data is that we observe the same movement along the whole chain of nodules. It means that hypothetic perturbations caused by AFM scanning to the *Nostoc* are minimal. This conclusion is enforced when variation of the positions of the "head" of the *Nostoc*, as measured by AFM (figure 6.b, stars) or optical microscopy (figure 6.b, solid line), are compared. There is a remarkable correlation between the first three AFM images. A small discrepancy is visible for the following images (labeled *4* and *5*) we can explain by the following reason. During the scanning of AFM image #3 (between the short vertical segments labeled *3* and *4* in figure 6) the bacterium is accelerating to reach a speed of around 500nm/s. Thus the head nodule of the bacterium rapidly approaches the left border of the AFM image (image *4* in figure 7) and may glide beyond this AFM scanning frame. So the nodule we checked at the extreme far side in image *4* (figure 7) may not correspond to the head nodule of the bacterium but to the second one. The distance between these two nodules is 4 microns, the typical value of a nodule length. This distance perfectly fits with the deduced offset, around 4 microns, (figure 6.b) from the vertical shift between the optical data (solid line in figure 6.b) and AFM one (star labeled *4* in figure 6.b). The same kind of explanation well fits for the last AFM image (image *5* in figure 7) as a second nodule passed through. These arguments further enhance the idea that no major perturbation of the natural gliding of the *Nostoc* is caused by the AFM tip movement during scanning. From these optical and AFM data a mean gliding speed of 80±10nm/s is measured for the *Nostoc* in this first sequence.

We will now focus on effective stiffness data. Four examples are given in figures 8-9 where height and effective stiffness profiles are plotted, along the vertical dash or dash/dot lines, either for different positions of the profile in a same AFM picture or at different positions of the bacterium on the sample (figures 8 and 9). The plateaus at both sides of these stiffness profiles correspond to the hard surface, the glass slide. In figure 8.d, the mechanical profile over the bacterium reveals a central part with an effective stiffness of around 0.22N/m and edges about one order of magnitude lower, around 0.06N/m. These lateral zones with a low value for stiffness are related to the small ribs visible in the height profile (see blue circles in figures 8.f) with a thickness of few hundreds of nanometers. These zones may very likely be related to the presence of a polymeric ECM layer. Furthermore the stiffness value of that ECM is slightly lower in the concavity of the nodules chain than on the other side (60±5mN/m instead of 70±5mN/m). Same kinds of features are observed along the profile near the head extremity of the gliding bacterium (figures 8.c and 8.e). As previously the stiffness at the concave edge of the bacterium (ECM) is lower (70±5mN/m instead of 90±5mN/m for the convex side) but this effect is more pronounced. In case of a lower curvature of the *Nostoc* (figure 9) the dissymmetry for the stiffness values of the ECM between both sides of the bacterium is small: 90±5mN/m in the concavity to be compared to 85±5mN/m on the other side. It looks like the measured lateral dissymmetry in stiffness is related to the concavity along the bacterium: greater the local curvature is, higher is this stiffness dissymmetry.

Another important feature we observe (figures 8.a and 9.b) is the modulation of the effective stiffness along the longitudinal axis of the bacterium as zones with high stiffness, around 0.2N/m. The spatial period of these high stiffness zones is estimated to 20μm, the equivalent of roughly 5 nodules. The zones with low stiffness value predominate as the mean value of the effective stiffness for the bacterium as obtained from the histograms in figures 8.g and 9.i is equal to 70±5mN/m.

When contrast in both height (figure 9.c) and stiffness data (figure 9.d) is enhanced, minute details ahead of the *Nostoc* bacterium appear: it is clearly visible that the bacterium left a trace during its former backward and forward movements. Profiles along the black vertical dash/dot line (figure 9.a-d) are plotted in figures 9.e and 9.g. A polymeric deposit with a typical height in the range of few nanometers is evidenced. The slime thickness is higher on both sides (around 6nm) than in its central part (≈3nm). Stiffness histogram plot (figure 9.j), as calculated on the portion of figures 9.a-d without the bacterium (at the left side of the dotted line in figures 9.a-d), reveals the presence of two peaks: that for the hard substrate and the other (0.30 ±0.07N/m) for the slime. It must be pointed out that this value for slime stiffness is likely overestimated: as the slime layer is thin (≈3nm), the partial indentation of the substrate cannot be excluded; so we measured a mixed contribution of the slime itself and the glass substrate. Nevertheless the presence of that slime is clearly evidenced in the stiffness profile (see figure 9.g). Furthermore stiffness value at lateral concave part of the slime is lower (0.26±0.01N/m) than in its convex part (0.29±0.01N/m). The two sides of the slime track have very near heights: 5.3±0.1nm and 6.1±0.1nm at the left and right sides of the track respectively. If we hypothesize that the intrinsic slime stiffness is constant, small variations of the slime height will cause important changes in the resulting stiffness values: indeed, as thinner is the slime layer, higher the influence of the hard substrate and, correlatively, the resulting stiffness value would be. However we can note that in the present case (figures 9.e and 9.g) the thinner slime deposit (left side in profile 9.e) is related to a *lower* value of stiffness. We can thus conclude that there is a slight but detectable difference in the elastic properties of the slime between its both sides: slime stiffness is very likely lower in the concavity (left side in profile 9.g), a similar result to that noted for ECM. From these results, we can conclude that the more the concavity along the bacterium is marked, the greater the lateral dissymmetry in stiffness for both slime and ECM likely is.

The second set of data about *Nostoc* bacteria is now studied. We first observe optically that, in this case too, the gliding of the *Nostoc* bacteria along the surface of the glass slide is parallel to the symmetry axis of these filamentous cyanobacteria. An optical estimation of the mean speed along the vertical axis of the image led to a value of 200±20nm/s. Successive AFM images of the *Nostoc* bacterium are presented in figures 10.a-c (force set point was set at 9.4nN; scanning time for one picture is 79s). They show that the bacterium is moving from the bottom to the top of the image. Height images in figures 10.a, 10.b and in the upper part of figure 10.c (above the wide white line) are displayed with the same contrast. The characteristic shape of connected elementary nodules with a mean length of around 4 microns and a mean height of 2 microns (see height profile in figure 12.b0) is clearly visible in figures 10.a and 10.b. One remarkable feature must be pointed out: between the images 10.a and 10.b, a vertical shift of the main features, along the symmetry axis of the cyanobacterium, is evidenced and illustrated in figure 10.d. In this artificially recomposed picture, the height data from AFM picture 10.a was numerically up-shifted by 19.3±0.2µm (white arrow) and superimposed over figure 10.b. The correlation between both images is good as main features of the *Nostoc* are well recognizable in two AFM pictures. It is thus possible to estimate the gliding speed by AFM: we get a value of 245±25nm/s, very close to that obtained using optical evaluation (200±40nm/s). In figure 10.c (top part; same height scale as in figures 10.a and 10.b), there is no more bacterium in the AFM image as the cell glided away from the AFM scanned zone. An estimation of the *minimum* gliding speed between images 10.b and 10.c leads to a value of 260±25nm/s. As seen in figure 10.c (lower part) where an enhanced color scale (range = 30nm) was used, the passing-away of the *Nostoc* on the glass sample due to its gliding left a deposit of a slime layer: its mean thickness is about 15nm (figure 12.s1) with a slightly higher value in the concavity of the bacterium curvature. Corresponding stiffness images are plotted in figure 11 with a common stiffness range in their upper parts and with adapted enhanced contrasts in their respective lower parts (below the wide black lines). These data corroborate our observation made earlier: a thin layer of a soft material, the slime, has been left by the cyanobacterium on the glass during its gliding movement. The mean slime stiffness is around 0.22±0.01N/m as seen in profile (figure 12.s2) made along blue line in figure 11.c and in histogram plot in figure 11.d (crosses). This value is similar to that measured in the former set of data (sequence #1). We note too that the slime stiffness (figure 12.s2) is slightly smaller in the concavity of the *Nostoc* curvature: this effect is very likely not due to a spurious effect of the underlying substrate as the slime thickness (figure 12.s1) is not negligible (around 15nm) and varies only slightly along the profile. As seen in the profile plotted in figure 12.b2, the stiffness of this fast gliding bacterium has a higher mean value (around 0.2N/m) and a much lower standard deviation than in the formerly described slow gliding regime. This is confirmed by the histogram in figure 11.d (solid line): the peak corresponding to the mean stiffness of this fast bacterium (220±20nm/s instead of 80±10nm/s for sequence #1) is surprisingly of a much higher value than that determined for sequence #1: 0.18±0.01N/m instead of 70±5mN/m. A spurious effect due to the underlying substrate can be ruled out as the mean thickness of the bacterium (figure 12.b0) is in the micron range and thus is much higher than the typical indentation depth (few tens of nanometers). Our data reveal that the increase of the gliding speed of the *Nostoc* upon the glass slide may be related to an enhancement of its stiffness. Such correlations between mechanical properties and motility of cells were already shown in literature. For instance, the migration speed of cell increases with the hardness of the substrate [42]. Here again the stiffness profile (figure 12.b2) reveals slightly lower values at the border of the bacterium but the contrast between the stiffer central zone and the ECM is far less prominent than for sequence 1.

A last sequence of data about *Nostoc* bacteria is presented (figures 13). The main direction of movement (first image: figure 13.a; second one: figure 13.b) of the bacterium (from top to bottom of AFM image) is opposite to slow scanning direction of AFM tip. From the AFM data the mean speed of the *Nostoc* bacterium is evaluated to 120±10nm/s. Two consecutive stiffness images, separated by 137 seconds, are presented in figures 13.a-b. Slime left by the bacterium during its former movement is clearly visible (figures 13.a and 13.b) by the curved track in light gray. The *Nostoc* is shown to change its direction of propagation by leaving its former track and going straight down the AFM image. We can notice (figure 13.a) that no slime is secreted ahead of the bacterium. It confirms former results of Yu and Kaiser [43], where emission of slime was only visible at one pole of *Myxococcus xanthus*, at the end of the cell. Height (figure 13.c) and stiffness (figure 13.d) profiles on the bacterium along solid (figure 13.a) and dashed lines (figure 13.b) have been plotted. Both sides of bacterium, where ECM is likely present, still reveal smaller values of stiffness (50±5mN/m) in a similar way to sequence #1. However no clear dissymmetry between both sides is visible here. Profiles at the lower part of AFM images, where slime track and bacterium are spatially separated, are plotted in figures 13.e and 13.f. The stiffness of the bacterium (figure 13.f) near its head nodule reaches values as low as 30±5mN/m and does not present dissymmetry as observed in the former cases. This may be correlated to the straight movement of the *Nostoc*. As before, harder points are still present along the bacterium: their stiffness values approximate 0.2N/m. For this studied sequence, stiffness of the slime track has a high value around 0.23N/m (figure 13.f) for a mean value of thickness of 8nm (figure 13.e). As observed earlier, the slime layer is slightly thicker at the concave side. The presence of two distinct levels for stiffness for the bacterium and its slime is confirmed by histogram plots (figure 13.g) of stiffness data from figures 13.a-b. The peaks at 0.25±0.01N/m related to the two successive images 13.a and 13.b have almost the same amplitude. Thus they correspond to the slime zone as this one does not significantly change in area (same number of pixels in images) during the movement of bacterium as evidenced in both images 13.a and 13.b. Instead, the peak at 0.07±0.01N/m (figure 13.g) can be related to the bacterium as it intensity increases when the *Nostoc* glides down the AFM scan zone. That value, averaged over the whole length of the *Nostoc* including the high stiffness zones and ECM, is low as already observed for the other example of low gliding speed (sequence #1). It must be emphasized that good reproducibility of our AFM measurements both for height and stiffness is clearly evidenced in figures 13.e and 13.f on slime signals, i.e. for the left part of the profiles.

DISCUSSION

Thanks to our improved combination of an easy-to-apply and gentle sample preparation procedure and an efficient AFM method (QI mode, JPK) –see Material and Methods section for more details- we were able to visualize and mechanically characterize - in excellent conditions - very different bacterial strains in their living state. These experiments were done with bacteria in their respective physiological liquid media on standard glass samples. The most tremendous example is the AFM imaging of *Nostoc* cyanobacterium, at an unprecedented spatial resolution, during its natural gliding movement. The study of cyanobacteria indeed showed, for the first time, that it is possible to perform AFM images of high quality on a motile organism gliding over the surface of a glass slide at speed as high as 250nm/s, *i.e.* up to 900µm/h. This was possible because (i) *no* immobilization step of the bacteria, based on chemical or mechanical entrapment was needed and (ii) the used AFM mode does minimize lateral interactions between the AFM tip and the substrate during the successive, well-controlled and high speed approach/retract curves. This method can be applied to many types of bacteria: Gram-positive and Gram-negative as well, bacteria with very different types of shape and surface behavior. These bacteria can be studied in purely native conditions without any external stress. The characteristics of gliding movement of the *Nostoc* as deduced from both optical and AFM data were found to be identical. Furthermore no correlation, and consequently no disturbance, between the AFM tip movement and *Nostoc* gliding was observed. These two last points enforce the conclusion of no major perturbation of the natural bacteria gliding by the scan of the AFM tip upon the cell during imaging. The bacterium is thus staying in its standard living state. AFM study of the Gram-positive *Rhodococcus wratislaviensis* does not reveal any detectable movement of the bacterium. We thus obtained highly spatially resolved AFM images of *R. wratislaviensis* in native conditions for standard scan times. Minute details on their membrane were revealed by both topographic and mechanical stiffness data. For instance, small protuberances on the *R. wratislaviensis* membrane are observed in height images and correspond to zones with slightly lower stiffness. Furthermore height and stiffness images revealed the presence of low swellings (thickness and lateral extension in the range of few hundreds of nanometers) of a soft polymeric slime layer on one (right) side of the bacteria. This could be due to the deposition of a slime layer by *R. wratislaviensis* in an initial slight migration upon the surface followed by a self-immobilization process. This process would likely have occurred during the sample preparation before AFM experiments. The distance along which the cell slightly moves laterally before immobilizing could be tentatively estimated to 450nm, the lateral extension of this slime compound. We think that this polymeric layer is likely involved in the adhesion of the bacterium to the glass slide and in the connections to neighboring bacteria, which are the early steps for the completion of a biofilm. Our new procedure is an important step forwards to detailed studies of the immobilization or motility of bacteria on surfaces and the related formation of biofilms as their genuine trends have not been artificially blocked by the observer.

Stiffness profiles and histograms made from the whole stiffness images brought important and complementary information to height images. We were thus able to distinguish between two or three main components for the bacterial complex depending on the bacterial strain (see table 1). In case of *R. wratislaviensis* we already mentioned the presence of a slime compound around the bacteria characterized by a low cellular spring constant, $k_c$, of 0.18±0.07N/m. Stiffness of the pure bacterial part is higher, 0.64±0.15N/m. We did not detect the presence of an ECM layer around the bacterium. Slight variations of the local mechanical properties of *R. wratislaviensis* were observed and related to topographic features. The situation is more complex for the *Nostoc* cyanobacteria. Three main components with various mechanical characteristics were revealed by our experiments: (i) the bacterium with zones of higher stiffness values (up to 0.64 ±0.15N/m) located in its central part, (ii) the ECM compound, mainly situated along the edges of the bacterium, with a stiffness varying from 0.09 ±0.03N/m for low gliding speeds (below 150nm/s) to 0.29 ±0.08N/m for speeds higher than 200nm/s and (iii) a track of slime left by the *Nostoc* behind it during its gliding movement. Along the longitudinal axis of the *Nostoc,* we observed the presence of delimited zones of high stiffness with a spatial periodicity not related to that of the nodules: only few of them are in a state of high stiffness. The period is estimated to 20µm, which corresponds to roughly 5 nodules. The longitudinal position of these high stiffness regions along the chain of nodules, relatively to the head extremity of the *Nostoc* (figures 8.a and 9.b), seems to be preserved during the gliding. That is why a hypothetic explanation based on high-stiffness areas associated to *mobile* bacterial nucleoids underneath the bacterial membrane as in *E. coli* bacteria [44] can be likely ruled out. In case of a fast gliding *Nostoc* (speed higher than 200nm/s) the mean value of stiffness averaged on the whole bacterium is much higher (around 0.36 ±0.10N/m) than for slow cyanobacteria (<150nm/s). These data may reveal a hardening process probably related to the fact that the gliding speed of the bacterium on the glass slide is noticeably greater.

To summarize our results on mechanical properties of the two bacterial strains, it can be written that the typical values for cellular spring constant of the bacterium we measured are in the range of few $10^{-1}$N/m. Many AFM studies on the determination of local mechanical properties, as the cellular spring constant, have been already done on different bacteria in many experimental conditions. A rather complete synthesis about these mechanical properties of cells (mainly bacteria), mostly artificially immobilized on substrates, is proposed in reference [45]: $k_c$ typically varies between $10^{-2}$N/m and $10^{-1}$N/m. It must be noted that most of these results are in the range of $10^{-2}$N/m, one order of magnitude lower than data presented in this paper. More experimental results can be found in [46–48]. One study [49] deals with Gram-negative *Shewanella putrefaciens* in potassium nitrate solution with a fixed ionic strength (0.1 M) and variable pH between 4 and 10: the conditions of immobilization of the bacteria on the substrate were similar to ours as no mechanical entrapment or chemical process were used, the main difference laying on a pretty harsh rinsing procedure. It yielded again to bacterial spring constants between 0.02 and 0.05 N/m. Nevertheless it must be emphasized that mechanical properties (cellular spring constant, Young modulus) of the cell surfaces are extremely sensitive to the surrounding environment or to special treatment [50]. In case of applied damage to the cell, the main observed trend is that the cell wall becomes softer as for Gram-positive *Staphylococcus aureus* (after digestion by lysostaphin) [51] or for Gram-negative *E. coli* as predated by *Bdellovibrio bacteriovorus* [52]. In that last case the cell spring decreases from 0.23 N/m for healthy cells to 0.064 N/m for invaded bacteria. As we proved the bacteria that we studied are fully alive (as they glide), these last remarks likely explain the high values of spring constant (~0.1N/m) we measured when compared to most of the available results: artificially (chemically or mechanically) immobilized bacteria may have been damaged leading to spurious low values for stiffness. However more studies are needed as it was shown that, in few cases, *E. coli* cells can become stiffer when damaged by a heating shock [50].

An important feature was observed for *Nostoc* bacterium. During its gliding movement, it left a track of *slime* we directly evidenced by AFM height and stiffness data. Its stiffness was evaluated to approximately 0.2N/m, much higher than ECM one, and does vary so much with the gliding speed. When the gliding speed increases, we note the presence of a *stiffer* ECM layer at the edges of the bacterium and the expulsion of a *thicker* slime layer (figure 14). Figures 9.e and 13.e related to low gliding speeds show that the slime thickness is higher on both edges of the track left by the *Nostoc* during its gliding movement. This effect is not visible for high speed sequence, probably because of an increased smearing of this thicker polymeric slime layer. As the ECM was mainly detected along the edges of the bacterium, the production of the slime could tentatively be related to the presence of the ECM along the bacterium. Production of ECM and slime might be connected. The higher value for the slime stiffness, when compared to that of ECM, could probably be due to a temporal change of the properties of the polymeric ECM layer secreted by the *Nostoc* during its gliding movement, an aging-like effect of the initial polymer.

From a careful look at stiffness profiles perpendicularly to longitudinal axis of *Nostoc*, a questioning observation has been done. Stiffness values of the ECM are not symmetrically distributed along the two edges of the bacteria where ECM is mainly located. It must be noted that similar decrease of $k_c$ with increasing distance from the bacterium apex was already observed for Gram-negative bacteria *Shewanella putrefaciens* [53] and the yeast *Saccharomyces cerevisiae* [45]. According to our data, lower values for ECM stiffness are located at the concave edge and this effect is more pronounced for parts of the nodules chain with higher concavity. We also note that the stiffness of the slime is systematically lower on the high concavity side in the three sequences. Does the motility machinery of the cell excrete a thicker and softer compound at the highest concavity edge in order to increase the friction coefficient and provoke the curved movement? New experiments are scheduled to answer this question.

Important mechanical parameter, as turgor pressure can be derived from the effective stiffness data (see Material and Methods section for details). Data available in literature are rather scarce and mainly obtained from investigations at low spatial resolution (unless a recent exception [54]). As a first step, we limited our study of turgor pressure to data averaged over the whole bacterium as deduced from the histogram plots in order to compare these results with data in literature. A detailed study of the spatially distribution of turgor pressure over the bacterium is under work. The mean values of turgor pressures for *Nostoc* and *R. wratislaviensis* bacteria in their growth medium are reported in table 2. For the Gram-negative *Nostoc*, our experiments show that the turgor pressure depends noticeably on the mean gliding speed: at low speeds (below 100nm/s) we got an averaged value of 50±10kPa, while it reaches 180±30kPa in case of gliding speeds as high as 220nm/s. The turgor pressure for low speed *Nostoc* is near from values reported in literature for artificially immobilized bacteria. Literature indeed reveals that turgor pressure for Gram-negative bacteria, in case of middle to high ionic strength medium as the growth medium used in our experiments, ranges from 10 to 40kPa for *Pseudomonas aeruginosa* [7], *Shewanella putrefaciens* or *oneidensis* [55] and *Escherichia coli* [56]. These values are near to that we obtained. Measurements in distilled water or in any other medium with low ionic strength lead to values as high as 85-150kPa as, for instance, for *Magnetospirillum gryphiswaldense* species [40] or *E. coli* [54]. That is one order of magnitude higher than for growth media with high ionic strength as expected from [8]. Thus there is a rather reasonable agreement between our results and those of literature. To our knowledge and as already mentioned, no data are available for living and moving bacteria as those in the present paper.

Data about Gram-positive bacteria are scarcer and reveal important dispersion. Determination of turgor pressure for *Bacillus-subtilis* was performed in its growth medium and led to a value of 1.9 MPa [57]. For *Enterococcus hirae* [8], experiments were conducted in distilled water and the measured turgor pressure was in 400-600kPa range. This could be extrapolated to values around 50kPa in growth medium with high ionic strength. The value we measured for *R. wratislaviensis*, 307±20kPa, is intermediate between these two min-max values found in literature.

Independently of the turgor pressure measurement, the Young moduli related to the bacterial envelope and components (figure 15) were calculated from the stiffness data (see Material and Methods section for more details). Resulting values of the Young modulus are shown in table 3. In the case of *Nostoc* (figures 15.a1-a3 and 15.b1-b3) the *bacterium* Young modulus only concerns the stiffest parts of the bacterium as the softest part have mechanical properties very similar to those of the surrounding ECM. It must be noted that Young modulus of the different *Nostoc* components noticeably increases between the slow gliding regime (sequences #1 and #3) and the high speed one. Literature reveals, even if studies are rather sparse, that the value for the elastic modulus of biofilms and isolated bacteria varies greatly, depending on the method used to measure it and the way samples were prepared and studied [18,50]. Furthermore, mechanical properties of bacterial biofilms are very dependent on the ratio between ECM and bacteria. So the comparison with results in literature is not an easy task as the tabulated values are highly dispersed. It must be mentioned that bacteria are often studied in conditions very distant from their genuine physiological ones. As an example, Young modulus values as low as 7kPa were reported [58] on the biofilms of Gram-negative *Pseudomonas aeruginosa* at the interface of an agar surface and humid air. More typical values of the bacteria elasticity as determined by microscopic methods are rather in the range 5-50MPa [56,59–61] as in references [56] (23±8 - 49±20MPa) and [44] (between 10 and 20MPa) for the Gram-negative *E. coli*. In a recent study [54], local measurements of Young modulus yielded to values around 1MPa for *E. coli* in a 1 mM KNO$_3$ electrolyte. However values as low as 3MPa for *E. coli* [50] and 0.04-0.2MPa for Gram-negative *Shewanella putrefaciens* [49] were observed. The discrepancy between these data can be explained by the strong dependence of Young modulus values on the living or dead state of the bacteria [50]. Our data reveal that the *Nostoc* Young modulus noticeably increases with the gliding speed: from 20±3MPa below 150nm/s up to 65±5MPa at speed near to 250nm/s. As our study first deals, to the best of our knowledge, with Young modulus of living bacteria in physiological conditions, comparison with literature is tricky. However our results are reasonably comparable with these of former studies of Gram-negative bacteria. For Gram-positive ones, literature is very short of data. Measurements of cell wall Young modulus for *Bacillus subtilis*, performed by macroscopic methods [62], gave values from 13GPa in dry environment to 30MPa in humid air (around 90% in relative humidity). Measurements of the elastic modulus for archae *Methanospirillum hungatei* [63] was performed on sheaths isolated from the cell and suspended in deionized water. They gave a Young's modulus of 20 to 40GPa. The elasticity of the complete cell boundary was obtained for the Gram-positive *Lactobacillus* [64]. The measurements were performed using AFM and the slope of the force-indentation curves gave values for bacterial stiffness of $1-2.10^{-2}$N/m. Our original measurements for in-situ and living Gram-positive *Rhodococcus wratislaviensis* gave a Young modulus value of 104±5MPa.

Thanks to our method, values for Young moduli for the different components related to the bacteria, such as the bacterium itself, the slime and the ECM in case of the gliding *Nostoc*, are now available. For *Nostoc*, values of $E_{young}$ for ECM and slime were shown to slightly increase with the gliding speed. For non-motile *Rhodococcus* and for *Nostoc* gliding at low speed as well, the slime layers have Young modulus values in the range of 1-4MPa. As the *Rhodococcus* slime thickness is high (250nm), there is no indentation of the underlying glass slide and, consequently, the measured $E_{young}$ value, around 4MPa, properly characterizes the compressive modulus of this slime. Instead, for *Nostoc*, as the slime thickness is much lower (less than 20nm), the Young modulus is very likely over-evaluated. We guess that, for *Nostoc* in low speed regime, a more realistic value of $E_{young}$ for the slime might be in the range of few hundreds of kPa. We can thus point out that Young modulus for the *Nostoc* slime in low speed regime is roughly of the same order of magnitude when compared to that measured with non-motile (or rapidly self-immobilized) *R. wratislaviensis.* When referring to the gliding properties of these two strains as revealed by our studies, we could wonder whether the difference between their displacement speeds upon the substrate (from almost zero for *R. wratislaviensis* to 100nm/s for *Nostoc* in slow regime) could not only be due to the excretion *rate* of slimes with similar compositions (from a mechanical point of view at least): slime thickness in the range of tens of nanometers for the motile *Nostoc*, up to few hundreds of nm for the self-immobilized *R. wratislaviensis*. We observed that $E_{young}$ value for slime strongly increases when the *Nostoc* enters the fast gliding regime: the Young modulus of the slime left by the cyanobacterium is multiplied by a factor of almost 20 when comparing slow and fast gliding regimes. It could be hypothesized that the chemical nature of the excreted polymer varies depending on the bacterium speed.

We observed that the distribution of the zones of high stiffness along the longitudinal axis of the *Nostoc* is not related to the spatial periodicity of the nodules. The *Nostoc* bacterium presents zones of high stiffness much more distant, around 20 microns (see figure 8.a), than the centers of nodules (mean length of a nodule approximates 4 microns). These observations could be explained by at least the following scenarios: (i) large zones of the bacterium are coated by such a thick layer of soft polymer (ECM) that the AFM tip does not indent the underlying bacterium. So that the bacterial membrane and its higher stiffness would appear only at a restricted spots and thus generate an important roughness upon the bacterium we did not observe; (ii) the *Nostoc* is moving as a caterpillar. The *Nostoc* would retract from the sample from place to place, leading to a disruption in mechanical contact with the glass slide and consequently causing a vanishing of stiffness; (iii) a mechanical wave with an alternation of high and low stiffness zones running through the filamentous bacterium during its gliding; (iv) the presence of mobile hard structures, as nucleoids, present underneath the membrane layer [44]. However as we did not measure noticeable variations of height along the nodules the two first hypotheses might be ruled out. The last one is not likely as our observations showed that the stiff structures are at a fixed position relatively to the bacterial membrane. Further investigations are in progress to elucidate that point.

Our observation on *Nostoc* about an increase of the slime thickness with the gliding speed is a point of importance as it can reveal important information about the mechanisms of gliding of living organisms. Several models for gliding of bacteria have been proposed in literature. Hoiczyk and Baumeister [31] concluded, from optical and electron microscopy observations, that the extrusion of slime comes from "junctional pore complexes" in the cell wall. They indeed correlated the rate of slime extrusion with the movement speed of the cyanobacterial filaments (*Phormidium uncinatum* and *Anabaena variabilis*) and concluded that gliding movements are directly caused by the secretion of slime. A similar conclusion was drawn by Yu et al. [43] from the study of *Myxococcus xanthus* in the case of type A motile gliding. Study of *M. xanthus,* as deposited on a specially treated sample and studied by optical microscopy coupled to an enhanced ellipsometry technics [35], led to a very different conclusion. It indeed highlighted a correlation between the amount of slime deposit and the time spent by a cell at a given position, suggesting that slime is deposited at constant rate underneath the cell body regardless of its gliding speed [35]. In case of *Nostoc,* as presented in this paper (figure 14), we instead observe a significant increase of the slime thickness with the *Nostoc* gliding speed. An alternate explanation based on a fortuitous dependence between the slime thickness and the total residence time of the bacterium at one place of the sample can be very likely ruled out:

a correlation between the measured speed and the local net (and unknown) residence time seems to be unlikely as we observed such a behavior for different bacteria with probably very different histories in terms of reverse movements. As a consequence the gliding process for the *Nostoc* is likely directly related to the slime extrusion process as mentioned in [31].

CONCLUSIONS

In summary, we presented results on AFM imaging of living bacteria in their genuine liquid environment and in true *in-vivo* conditions. This was feasible thanks to a non-perturbative, easy-to-apply method that avoids an external immobilization step unanimously described in literature as mandatory. This method offers an unprecedented lateral resolution and a quantization of roughness and stiffness at nanometer scales on living organisms in physiological conditions. Concerning the biological samples preparation, there is no need of mandatory and invasively drastic conditions such as fixation procedures, fluorescent-staining or enhancement of optical contrast by the use of specially designed substrate coatings as for electron or optical microscopy technics. This study was feasible thanks to the combined effects of a non-perturbing method for samples preparation and a new AFM mode [36] based on a very well controlled extend/approach curves made pixel by pixel and at a high speed. AFM parameters were optimized after several tests we made to get outstanding results on challenging samples. This AFM mode drastically minimized the lateral interactions between the cantilever and the biological organism. We studied Gram-positive, non-motile *R. wratislaviensis* strains and Gram-negative, gliding *Nostoc* cyanobacteria in their respective physiological liquid media.

On both bacterial strains height images of high quality were obtained by AFM in liquid. In particular we were able to follow the natural gliding movements of the *Nostoc* cyanobacteria both with AFM and optical imaging, directly proving the living state of the organisms during the AFM investigation and its very low impact on the biological state of the *Nostoc*. Our study revealed that AFM can image moving living species with gliding speeds as high as few hundreds of µm/h, which is unprecedented. These AFM images of gliding cyanobacteria were acquired at a high AFM acquisition rate, typically two AFM frames per minute, without limitation on spatial resolution. These breakthrough AFM observations strongly minimized perturbations on living cells when compared to common procedures based on mechanical entrapment or chemical immobilization.

Simultaneously with height data, mechanical properties of the two strains were acquired at nanometer scale. Values for stiffness, Young modulus or turgor pressure of these bacterial strains were obtained. These data revealed the presence of three main components on and around the bacteria: (i) the bacterium itself characterized by a Young modulus ranging from 104±5MPa for Gram-positive *R. wratislaviensis* to 20±3MPa for Gram-negative *Nostoc* at low gliding speed. In that last case the mean value of $E_{Young}$ was shown to increase with gliding speed. Furthermore our data revealed the presence of zones with high stiffness with a spatial periodicity of about 20 microns, i.e. approximately one *Nostoc* bacterial nodule out of four. For both strains the turgor pressure varies from 50±10 to 307±20kPa depending on the bacterium and its gliding speed; (ii) a thin layer of polymeric slime left by the bacterium during its gliding movement (*Nostoc*) or to promote its adhesion (*R. wratislaviensis*) on the surface of the glass sample. In that last case slime thickness was in the range of few hundreds of nanometers instead of few tens of nm for the gliding *Nostoc*. The Young modulus of the slime was as low as few MPa for low speed moving bacteria. We also note that the *Nostoc* slime systematically has a lower value of stiffness and larger thickness at the high concavity side: we hypothesize that it could cause a dissymmetric lateral increase of the friction coefficient and a consecutive curved gliding movement. For *Nostoc* our experiments showed that the slime layer thickness is increasing with the gliding speed reinforcing Hoiczyk *et al.* hypothesis [31] of a propulsion by ejection of slime; (iii) an extra cellular matrix (ECM), in case of *Nostoc* bacterium, with $E_{Young}$ in the range of 4-15±3MPa depending on the gliding speed. For *R. wratislaviensis* no ECM was detected in the present operating conditions.

An important conclusion of this study is that data with an excellent quality for both height and stiffness can be obtained by AFM in liquid without the need of an artificial immobilization step of the bacteria on the sample. This opens a wide window on new studies at nanometer scale based on the dynamics of living cells in purely controlled physiological conditions: motility or adhesion processes of bacteria on solid substrates, biofilms formation, influence of light on moving properties of cyanobacteria, AFM monitoring of a biological state, effects of antibiotics on bacterial membranes and biofilm etc.

MATERIAL AND METHODS

**Bacterial preparation**

The cyanobacterial strains used in this work were *Nostoc* strain, PCC 6720 (*Anabaenopsis circularis*) and were purchased from the Institut Pasteur Collection (CIP, Paris, France). Cyanobacteria are photoautotrophic Gram-negative prokaryotes. These strains were grown in autoclaved standard BG11 liquid medium prepared from 50x sterile-filtered concentrated cyanobacteria BG-11 freshwater solution (Sigma-Aldrich) and 18.2 MΩ.cm purified water (Milli-Q water purification system, EMD Millipore Corporation, USA). Before AFM experiments, cultures were incubated at 25°C at a constant incident flux of white light half the day in a dedicated chamber and were in contact with external air through anti-contamination filter. Bacteria were transplanted in fresh medium regularly.

The other studied strain is *Rhodococcus wratislaviensis*, capable of degrading multiple petroleum compounds in aqueous effluents and registered at the Collection Nationale de Cultures de Microorganismes (CNCM), Paris, France under number CNCM I-4088 (provided by IFPEN). Stock cultures were kept frozen at -80 °C in 20% glycerol (v/v). The culture medium used was a vitamin-supplemented mineral medium (MM). This medium contained $KH_2PO_4$, 1.40 g.l$^{-1}$; $K_2HPO_4$, 1.70 g.l$^{-1}$; $MgSO_4$ 7 $H_2O$, 0.5 g.l$^{-1}$; $NH_4NO_3$, 1.5 g.l$^{-1}$; $CaCl_2$ 2 $H_2O$, 0.04 g.l$^{-1}$; $FeSO_4$ 7 $H_2O$, 1 mg.l$^{-1}$. A vitamin solution and an oligo-element solution were added as previously described [65,66]. After inoculation (10%), the adequate carbon source was added, and the cultures were incubated at 30°C with constant agitation. Cultures were grown in flasks closed with a cap equipped with an internal Teflon septum to avoid any loss of substrate either by volatilization or by adsorption. The headspace volume was sufficient to prevent any $O_2$ limitation during growth. Growth was followed by measuring the Optical Density at a wavelength of 660nm. Bacteria were transplanted in fresh medium once a week as stated by the results of limiting oxygen and toxicity led by the IFP.

**Sample preparation**

The samples we used for the AFM experiments were standard glass substrates for optical microscopy. In a first step they were cleaned by sonication in a diluted solution of detergent (pH around 9) for 15 minutes before being carefully rinsed with high purity water (Milli-Q). Drying was done below the flux of a pure inert gas.

The bacterial suspension in its culture medium (BG 11 medium for *Nostoc* or MM medium for *Rhodococcus wratislaviensis)* was sonicated during three minutes then vortexed (two minutes) in gentle conditions. Fourty microliters (μL) were then deposited on the glass slide during the period $\tau_1$. The excess of solution was thereafter aspirated by a micropipette and the glass slide was further left in surrounding atmosphere (22°C and around 60% of relative humidity) for few ($\tau_2$) minutes before being rinsed twice with 500μL of pure water then twice with 500μL of the appropriate culture medium in gentle conditions. The glass slide was then placed at the bottom of the liquid cell, ECCell® from JPK [36] and 500μL of the corresponding medium were promptly poured in the liquid cell. The final bacterial surface concentration on the glass substrate for the AFM experiments was around 100 and 2.10$^3$ units per

mm$^2$ in case of *Nostoc* or *R. wratislaviensis* respectively, as checked by optical and AFM microscopies. $\tau_1$($\tau_2$ respectively) was in the range of 15 minutes (5mn respectively). The crucial step for AFM imaging in QI mode without any "external" immobilization process corresponds to the period $\tau_3$ during which bacteria are in a special intermediate "wet-dry" state we visually checked: as soon as the dehydration front is running through the bacteria, the sample is immediately rehydrated. $\tau_3$ is in the range of few seconds for both studied strains. As we did not check the bacterial concentration before the aspiration/refilling step, the "yield" of the self-immobilization of bacteria on the sample is unknown. To compensate the natural evaporation of the medium, we continuously supplied the ECCell in liquid medium at a rate of 100mm$^3$/h by a standard syringe pump. It must be emphasized that no turbulence effects were detected. AFM measurements were made in the two hours after inoculation of the glass plate. No spontaneous detachment of bacteria from the sample towards the planktonic phase was evidenced by optical or AFM microscopy. AFM experiments were also done when liquid medium was left in a drop-form on a standard glass slide as well and similar AFM results were obtained. We think that the crucial step for the self-"immobilization" (with and without gliding according to the strains) is the time $\tau_3$. More detailed studies about its role in the first steps of the biofilm building are under work.

**Optical and AFM imaging**

Atomic force microscopy studies were carried out at a temperature of 27±1°C using a Nanowizard III (JPK Instruments AG, Berlin, Germany). Experiments were operated in liquid (BG11 medium and mineral medium for *Nostoc* and *R. wratislaviensis* respectively), using Quantitative Imaging® (QI) mode. Qi is a force curve based imaging mode. Its main characteristic is to measure a real and complete force distance curve, at a defined constant velocity, for every pixel of the image. Vertical forces are precisely and continuously controlled during the whole approach and retract steps while imaging. Thus we got really quantitative measurements. In this mode, lateral forces applied by the apex of the AFM tip on the studied object are minimal (no pushing away or moving around of sample features). We optimized different parameters controlling this mode and, after several tests, the pixel-by-pixel extend/retract curves were done at a constant speed in the range of typically 50-500µm/s on a total extension of 500nm. It corresponds to a typical indentation speed of 17-175µN/s. An additional retract length of 100nm was added before going to the next pixel. In case of motile bacteria as the *Nostoc* studied in this paper, we could wonder if this fast approach-retract AFM method is relevant. As a matter of fact in our AFM experiments the indentation-like approach lasts 1ms for a total height variation of the tip of 500nm. At every pixel, the relevant portion of the approach curve, where the AFM tip starts to be in interaction with the substrate, is in the range of 100nm. Thus, in case of a typical gliding speed of 200nm/s, the bacterium moves of only 40pm during the acquisition of AFM data for one pixel. Consequently the bacterium can be considered as immobile for every pixel of the image.

Typical images were done on the basis of a surface scanning with 128 by 128 pixels. The distance between two successive pixels along the sample surface is usually different along slow scanning axis (vertical in the reported AFM pictures) and fast scanning axis of the images. We used standard beam AFM probes (PPP-CONTPt, Nanosensors, Neuchatel, Switzerland) with a nominal value of stiffness of 0.36±0.01N/m, as measured by thermal noise [41], and a tip height of about 15 microns. The cantilever is coated by a standard 25 nm thick double layer of chromium and platinum-iridium alloy on both sides. The maximum applied force was in the range of 5-10nN. Within this range, no major changes in the quality of AFM data were observed. No noticeable contamination of the apex of the tip was detectable (contrary to what happens when bacteria are immobilized by gelatin for example). A same cantilever was typically used few consecutive days for imaging bacteria without any noticeable deterioration.

AFM experiments with *Rhodococcus wratislaviensis* strain were replicated forty times by studying bacteria either at various locations of the same sample or for different samples built from different micro-pipetting in the available bacterial cultures. If we define a "success rate for AFM imaging" as the ratio between the number of trials for which the studied bacterium remains attached to the glass plate throughout the entire experiment and the total number of trials, then the success rate for *R. wratislaviensis* can be estimated to higher than 95%. For *Nostoc* this success rate is lower (around 50%) and may critically depend on $\tau_3$. When the bacteria remain on the glass plate, there were systematically gliding. As their movements upon the substrate are erratic, most of the time was spent to look for the bacterium and guess its further movements in order to place the AFM scanning window at the right position. Consequently the total number of replicates of AFM experiments for *Nostoc* was less (20) than for the non-motile bacteria. The examples detailed in this paper are typical of these studies. All the presented height AFM images are raw data (without any post-treatment as flattening, etc.). The stiffness data were calculated from the slope of the approach curves (force versus scanner elongation) at point of maximum force as averaged on a distance interval of 10nm (as sketched in figure 2) by custom Matlab programs.

The AFM head is working on a commercial inverted microscope (Axio Observer.Z1, Carl Zeiss, Göttingen, Germany). During all the presented experiments the sample was lighted by a LED illuminating system (white light) operating in transmission mode. The optical microscope was used in bright-field conditions without any staining procedures. The samples were first screened by a LD "Plan-Neofluar" 10x/NA0.3 objective (Carl Zeiss, Göttingen, Germany) and optical images, as shown in this paper, were taken through a LD "Plan-Neofluar" 40x/NA0.6 objective (Carl Zeiss, Göttingen, Germany) by a standard color camera. Image analyzing was made through custom computing codes developed with Matlab R2011 (The MathWorks Company, Natick, MA, USA).

**Measurement of turgor pressure**

First, by considering the AFM/tip mechanical system as the association of two linear springs in series (the cantilever and the studied sample itself) [40], we calculate the real stiffnesses of the bacterium and its components, the ECM and the slime, ( $k_b$, $k_{ECM}$ and $k_{slime}$ respectively) instead of their effective values as directly deduced from raw data. These data are reported in table 1: values for effective/real stiffness are in normal/*italic* characters respectively. Getting relevant values of mechanical parameters of the bacterium and its components from real stiffness is a hard task as the indentation stiffness of the bacterium wall is governed by several terms: these associated with stretching and bending of the cell wall, terms related to the surface tension and those directly related to the turgor pressure [40], the difference in pressures between the inner and the outer part of the bacterium as delimited by the cell membranes. In one recent case [56], independent measurements of elastic modulus of the cell wall and turgor pressure of *E. coli* were done by comparing results from intact and bulging cells as obtained by using likely aggressive antibiotic agents (kanamycin and vancomycin). Mostly, by applying some hypotheses [8,40], the stiffness data taken on the bacteria are interpreted as a measurement of the turgor pressure. Thus, we derived the value of turgor pressure from the simple model introduced by Yao et al. [8]. These authors approximate the turgor pressure from a model based on tension dominated concept for the deformation of bacterial envelopes. From Boulbitch [67] and Arnoldi *et al.* [40] calculations on the deformation of bacterial envelopes by an AFM tip, Yao *et al.* [8] reduced the considerable mathematical complexity of the problem of turgor pressure calculation by elucidating some of the components that contribute to the overall deformation. Thus the turgor pressure is derived from the averaged slope (*s*) of the high force regime obtained from force spectrum in an identical manner to what has been done in this paper for all the so-called effective stiffness data (see figure 2). The turgor pressure is calculated from the real stiffness of the bacterium, *s*, by using equation (13) in reference [8]. Following parameters were used: $R_b$, the effective radius of the bacterium, equal 1.8μm and $r_t$, the mean tip radius, $r_t$ = 50nm.

**Measurement of Young modulus**

The Young modulus, characterizing the elasticity of bacterial envelope, can be determined from the curve of the variation of the AFM applied force on the sample, at the low-force side, versus its indentation. This curve (see examples in figures 15.b) is obtained from the plots of the applied force during the AFM tip approach to the glass slide and the studied bacterium (figures 15.a). From the difference between this hard surface line (glass slide) and the observed deflection over the bacterium, the cell indentation is calculated. The force versus indentation curve can be analyzed through theoretical models for quantitative information on sample elasticity. In order to get an estimate of the Young modulus of the different components on and around the bacterium, we classically used the Hertz model [68]:

$$F = \frac{2E \tan \alpha}{\pi (1-\nu^2)} \delta^2$$

We took a Poisson coefficient, $\upsilon$, equal to 0.5 and a semi-top angle, $\alpha$, of the AFM tip equal to 35°.

Typical curves for the force versus piezo-displacement, for glass slide and for the different components of the bacterium, are plotted in figures 15.a. The related force versus indentation curves and the best fits using Hertz model are plotted in figures 15.b.

ACKNOWLEDGEMENTS

The authors would like to thank A. Hermsdoerfer and T. Henze (JPK Instruments AG, Berlin, Germany) for fruitful discussions. Dr. R. de Wit, Ecologie des Systèmes Marins Côtiers, UMR 5119, University Montpellier 2, France, and Dr. O. Brunel, Hydrosciences Montpellier, UMR 5569, University Montpellier 2, France are gratefully acknowledged for helpful discussions. A. Desoeuvre gave us an efficient technical help. The authors are very grateful to IFP Energies Nouvelles, Rueil-Malmaison, France (Dr. Françoise Fayolle-Guichard and Yves Benoit) for the free disposal of *Rhodococcus wratislaviensis,* IFP 2016 strain, through Dr. Marie-Christine Dictor and Jean-Christophe Gourry, BRGM, Orléans, France. This work is funded by the Agence Nationale de la Recherche (ANR, Paris, France) through the program ECOTECH_2011 (project BIOPHY N° ANR-10-ECOT-014-05).

FIGURE LEGENDS

**Figure 1. AFM height and stiffness data for *Rhodococcus wratislaviensis*.**
AFM height (a-b) and stiffness (e) images of *Rhodococcus wratislaviensis* in their physiological (MM) medium. Height (c-d) and stiffness (f-g) profiles along dashed lines in figures (a) or (e), respectively, are plotted.

**Figure 2. AFM approach curves.**
AFM approach curves in MM liquid are plotted for the glass substrate (black line), the *Rhodococcus wratislaviensis* bacterium (red line) and its slime (blue line). The raw curves have been shifted along $X$ axis to better view the contrast in slopes at the force setpoint. The dashed lines are an illustration of how the effective stiffnesses are calculated from these approach curves (best linear fit for a length window of 10nm).

**Figure 3. Stiffness histogram for *Rhodococcus wratislaviensis*.**
Stiffness histogram related to AFM stiffness image (figure 1.e) is shown.

**Figure 4. AFM height and stiffness data for *Rhodococcus wratislaviensis*.**
AFM height (a-b) and stiffness (c) images of *Rhodococcus wratislaviensis* in MM medium are shown. These images are taken at zone delimited by the white square in figure 1. Stiffness histogram related to AFM stiffness image (figure 4.c) is shown in figure 4.d.

**Figure 5. Optical snapshots of the gliding *Nostoc*.**
Optical snapshots of the gliding of *Nostoc* bacterium upon the glass slide are shown. The scale is given by the black line (10 microns). The bacterium is moving from the right to the left of the images as indicated by the displacement of the landmark (arrow) between the two images ($\Delta t=12s$).

**Figure 6. *Nostoc* displacement and speed curves as measured by optical microscopy.**
Figures 6.a-b: The displacement, along the $X$ axis (see figure 5), of a *Nostoc* bacterium, as determined by optical microscopy, is plotted (full line) versus time during its gliding movement on the surface of the glass slide. The AFM cantilever is 500nm far from the substrate.
Figure 6.b: The displacement along the $X$ axis of another *Nostoc* bacterium, as determined by optical microscopy, is plotted (full line) versus time. These data are related to the sequence labeled "number 1" in the main text. The AFM tip is now in contact with the substrate and scans it. The AFM fast scan direction is along $X$. The movement of the AFM cantilever along $Y$ (slow axis) is plotted versus time (black triangles; the line is a guide for the eye): every triangle corresponds to a measured position (one optical image every 4 seconds). The starting times of the successive AFM images are marked by the short vertical segments. The indexation number of the AFM images is labeled in the squared box. The successive $X$ positions of the bacterium as determined from AFM images (see figure

7) are marked by the stars (*), the dashed line being a guide for the eye. The scanning time for a full AFM image is 35 seconds.

In figure 6-c (upper curve with left triangles) gliding speed along X axis of *Nostoc* as calculated from to displacement data in figure 6.b is plotted *versus* time.

**Figure 7. AFM height images of *Nostoc* (sequence #1).**

Successive AFM height (1-5) images of *Nostoc* cyanobacterium in its physiological medium are plotted. Time interval between two consecutive images is equal to 35 seconds. For comparison, an optical image acquired during this AFM sequence was numerically treated to get the same magnification as in AFM images (1-5).

**Figure 8. AFM height and stiffness data for *Nostoc* (sequence #1)*.***

AFM stiffness (a) and height (b) images of *Nostoc* in its physiological (BG11) medium are shown. They correspond to image 7.5. Stiffness (c) and height (e) profiles along the black dash/dot line in figures (a) or (b) are plotted. Stiffness (d) and height (f) profiles along the red dashed line in figures (a) or (b) are plotted. The blue circles in figures 8.e and 8.f were drawn to point out the presence of ECM at the edges of the *Nostoc* profile.

**Figure 9. AFM height and stiffness data for *Nostoc* (sequence #1)*.***

AFM height (a, c) and stiffness (b, d) images of *Nostoc* in BG11 medium are shown. They correspond to image 7.3. In images (b) and (d), the contrasts were increased to show the presence of the slime layer. Height (e) and stiffness (g) profiles along the black dash/dot line in figures (a-d) are plotted. Height (f) and stiffness (h) profiles along the red dashed line in figures (a-d) are plotted. Stiffness histogram related to the whole stiffness image (figure 9.b or 9.d) is shown in figure (i). In figure (j), the histogram was calculated to the only portion of image 9.b (or 9.d) at the left side of the dotted line. The peaks resulting from the deconvolution of this histogram are plotted with colored lines.

**Figure 10. AFM height images of the gliding *Nostoc* (sequence #2)*.***

- Figure 10.a-c: Three AFM height images of the gliding *Nostoc* bacterium upon the glass slide. In figure 10.c. the bacteria glided away from the AFM scan zone. A common color scale for height was applied for all these images except for the lower part (below the thick white line) of picture (d) where height contrast was enhanced.
- Figure 10.d: Visualization of the vertical gliding movement of the *Nostoc*. It is done by the superposition of the image of the bacteria as determined in figure 10.b (bacterium at the right side of image 10.d) with that measured at $t_1$, 79s earlier (figure 10.a) and vertically shifted along the white arrow (shift length: 19.3±0.2 microns). For reasons of clarity a lateral shift between the native figures 10.a and 10.b was applied.

**Figure 11. AFM stiffness data of the gliding *Nostoc* (sequence #2)*.***

*Nostoc* AFM stiffness images are shown. They correspond to the three equivalent height images in figure 10.a-c. A common grey scale for stiffness was applied for the upper part (above the wide black line) of these images. At the lower parts stiffness contrast was enhanced. In figure 11.d stiffness histograms for *Nostoc* bacterium (solid line) and slime (crosses) are shown.

**Figure 12. AFM height and stiffness profiles for *Nostoc* (sequence #2).**

AFM height (_0, _1) and stiffness (_2) profiles for *Nostoc* bacterium (b_) along red line in images 10.b and 11.b and for the excreted slime (s_) along blue line in images 10.c and 11.c are plotted. Height profiles in figures _b0 and _s0 are displayed with the same height scale. Idem for those in figures _b1 and _s1 but with an enhanced contrast.

**Figure 13. AFM stiffness and height data for *Nostoc* (sequence #3).**

a; b : two successive AFM stiffness images of *Nostoc* are shown. Height (c) and stiffness (d) profiles along the black full line in image (a) and the black dashed line in image (b) are plotted with full and dashed lines respectively. Height (e) and stiffness (f) profiles along the red full line in image (a) and the red dashed line in image (b) are plotted with full and dashed lines respectively. Stiffness histograms related to AFM stiffness data from figure a (full line) and figure b (dots and dashed line) are shown in figure (g).

**Figure 14. Variation of the slime thickness with the *Nostoc* gliding speed.**

**Figure 15. AFM approach and indentation curves for *Nostoc* and *Rhodococcus wratislaviensis* bacteria, ECMs and slimes.**

Typical curves of force *versus* vertical piezo displacement (a_) and force *versus* indentation (b_) for *Nostoc* bacterium (_1), slime (_2), ECM (_3) and for *R. wratislaviensis* bacterium (_4), slime (_5) are plotted. Solid lines in figures (b_) are the best fits by applying Hertz model (see main text for details).

Table 1

Values of effective stiffness and cellular spring constant (in italics) for the two studied bacteria and their different components.

| Strain | Sequence number | Gliding speed (nm/s) | Figure numbers | Bacterium Stiffness (N/m) | | ECM Stiffness (N/m) | Slime Stiffness (N/m) | Thickness (nm) |
|---|---|---|---|---|---|---|---|---|
| | | | | Mean value from histogram | Stiffest points | | | |
| Nostoc | #1 | 80 ±10 | 8; 9; 10; 11 | 0.07 ±0.02 | 0.20 ±0.05 | 0.07 ±0.02 | 0.33 ±0.05 | 4 ±2 |
| | | | | *0.09 ±0.03* | *0.47 ±0.13* | *0. 09 ±0.03* | *4.0 ±1* | |
| | #2 | 220 ±20 | 14; 15; | 0.18 ±0.04 | 0.23 ±0.05 | 0.16 ±0.03 | 0.22 ±0.04 | 15 ±5 |
| | | | | *0.36 ±0.10* | *0.64 ±0.15* | *0.29 ±0.08* | *0.57 ±0.15* | |
| | #3 | 120 ±10 | 16; 17 | 0.07 ±0.02 | 0.20 ±0.05 | 0.03 ±0.01 | 0.25 ±0.04 | 8 ±3 |
| | | | | *0.09 ±0.03* | *0.45±0.13* | *0.04 ±0.01* | *0.82 ±0.20* | |
| R. wratislaviensis | | ≈ 0 | 3; 4 | 0.23 ±0.05 | 0.24 ±0.05 | / | 0.12 ±0.02 | 250 ±30 |
| | | | | *0.64 ±0.15* | *0.72 ±0.18* | / | *0.18 ±0.07* | |

Table 2

Turgor pressure for the studied bacteria at different gliding speeds.

| Strain | Sequence number | Gliding speed (nm/s) | Figure numbers | Turgor Pressure (kPa) |
|---|---|---|---|---|
| *Nostoc* | #1 | 80 ±10 | 8; 9; 10; 11 | 58 ±5 |
| | #2 | 220 ±20 | 14; 15; | 174 ±15 |
| | #3 | 120 ±10 | 16; 17 | 42 ±5 |
| *R. wratislaviensis* | | ≈0 | 3; 4 | 307 ±30 |

Table 3

Young modulus for the studied bacteria at different gliding speeds.

| Strain | Component | Gliding speed | Young modulus (MPa) |
|---|---|---|---|
| *Nostoc* | | | |
| | Bacterium | Low | 20 ±3 |
| | | *High* | *63 ±5* |
| | ECM | Low | 20 ±3 |
| | | *High* | *63 ±5* |
| | Slime | Low | 1.1 ±1.0 |
| | | *High* | *20 ±3* |
| *R. wratislaviensis* | | | |
| | Bacterium | | 104 ±5 |
| | Slime | | 4.3 ±.5 |

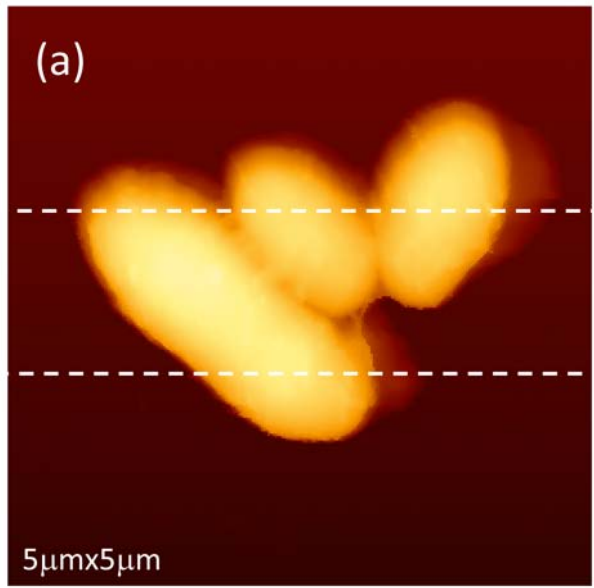
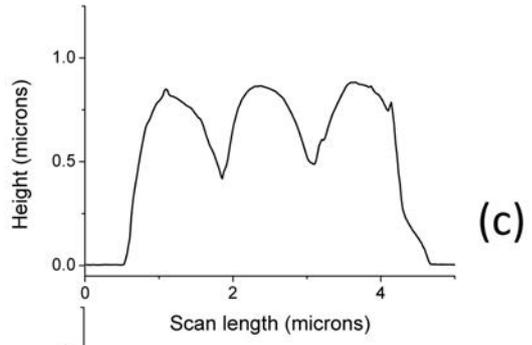
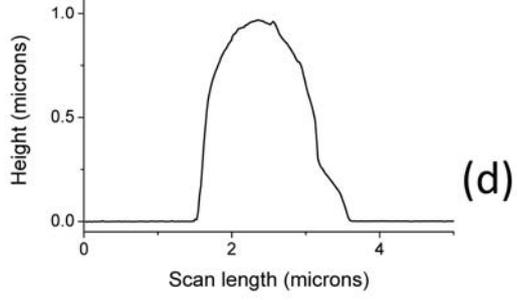
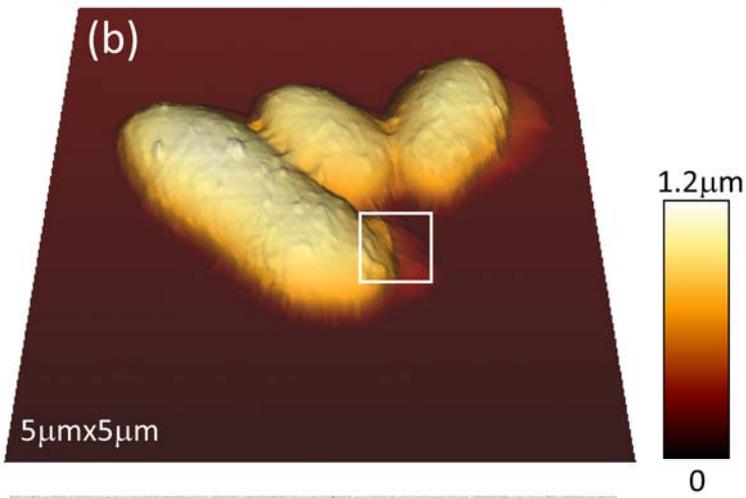
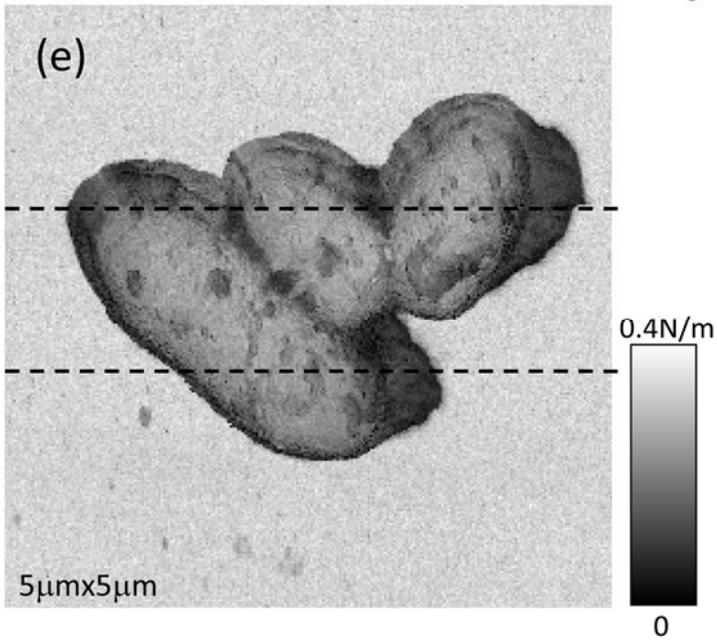
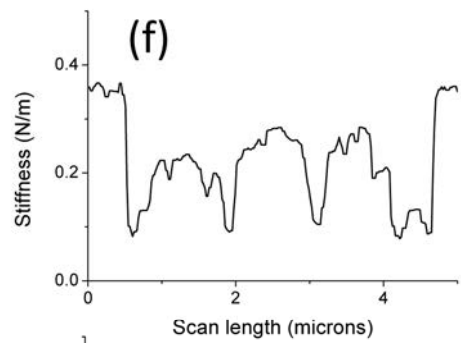
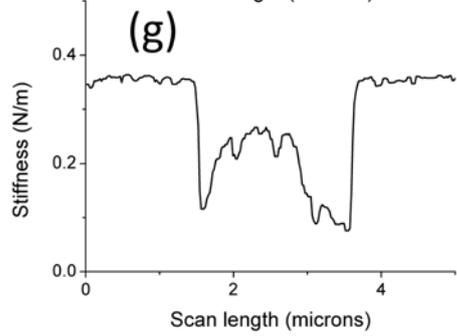

Figure 1 (version R2)

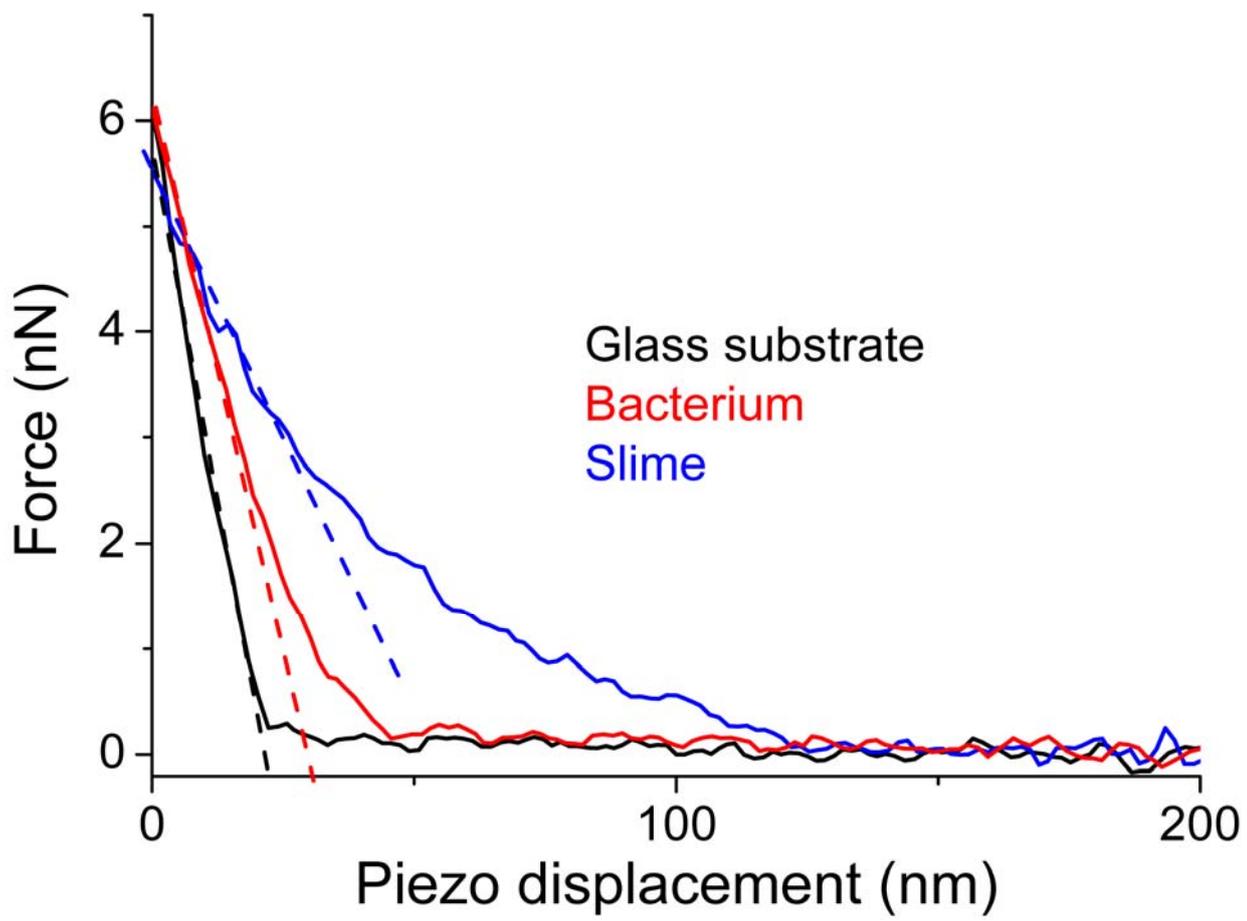

Figure 2 (version R2)

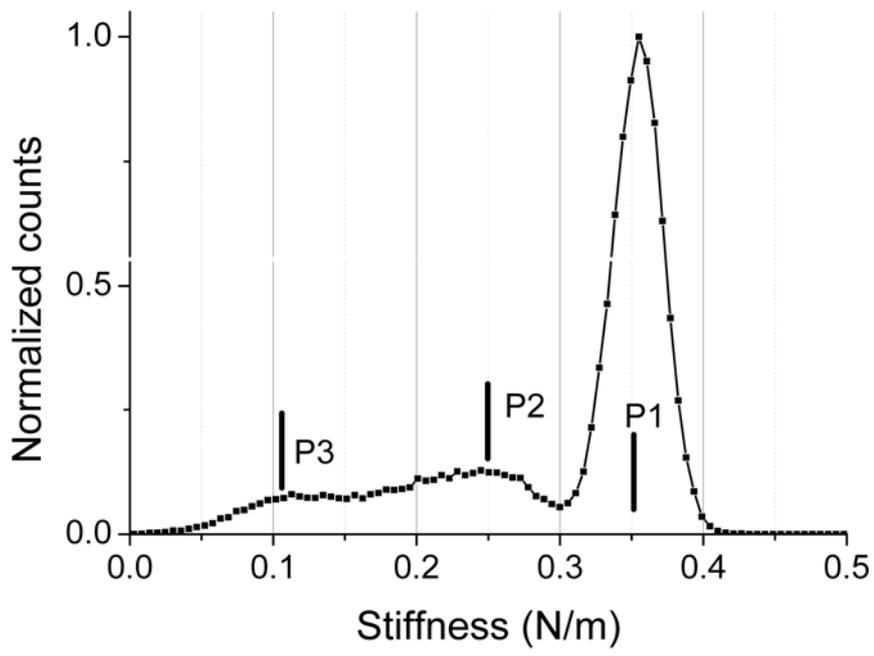

Figure 3 (version R2)

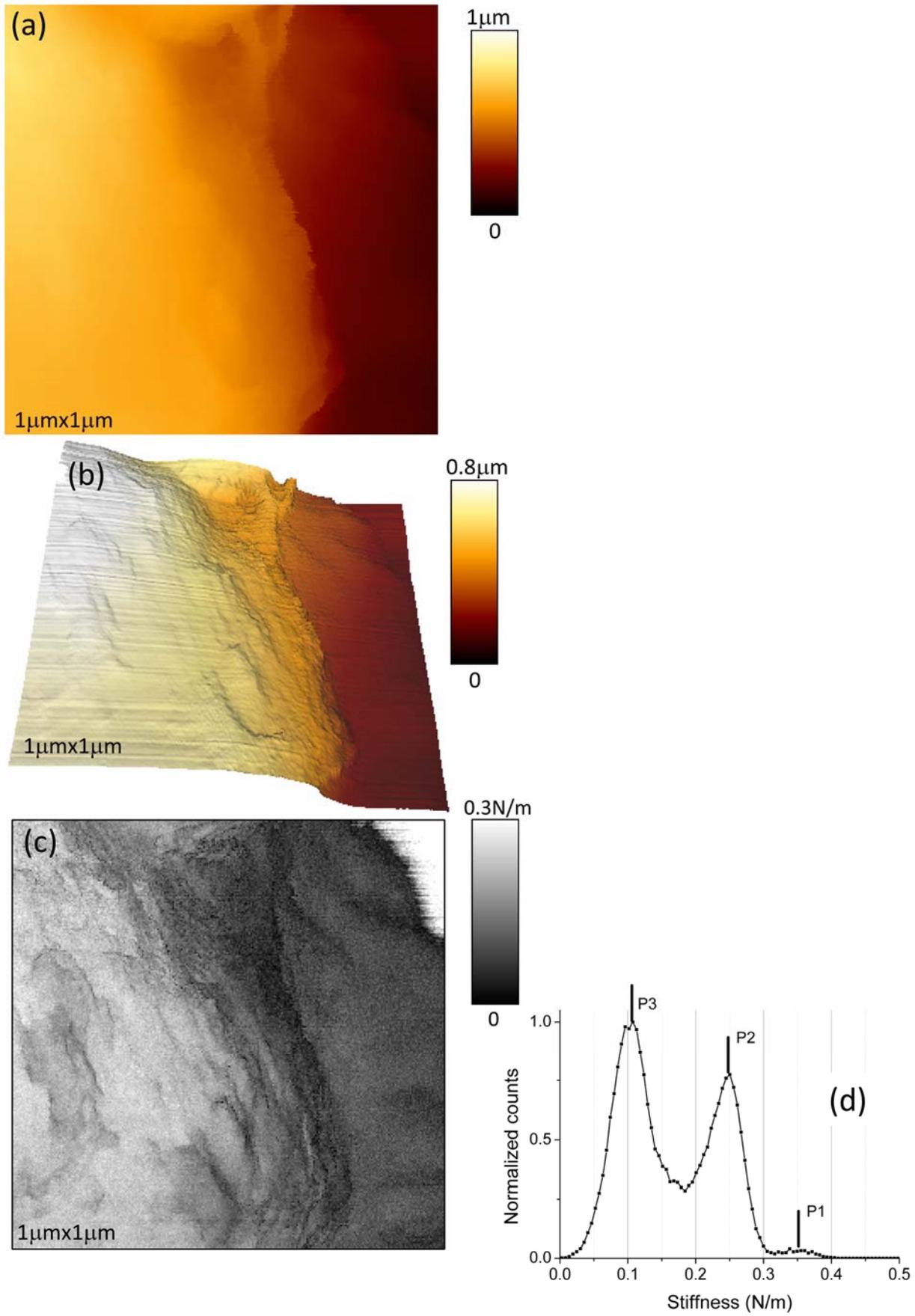

Figure 4 (version R2)

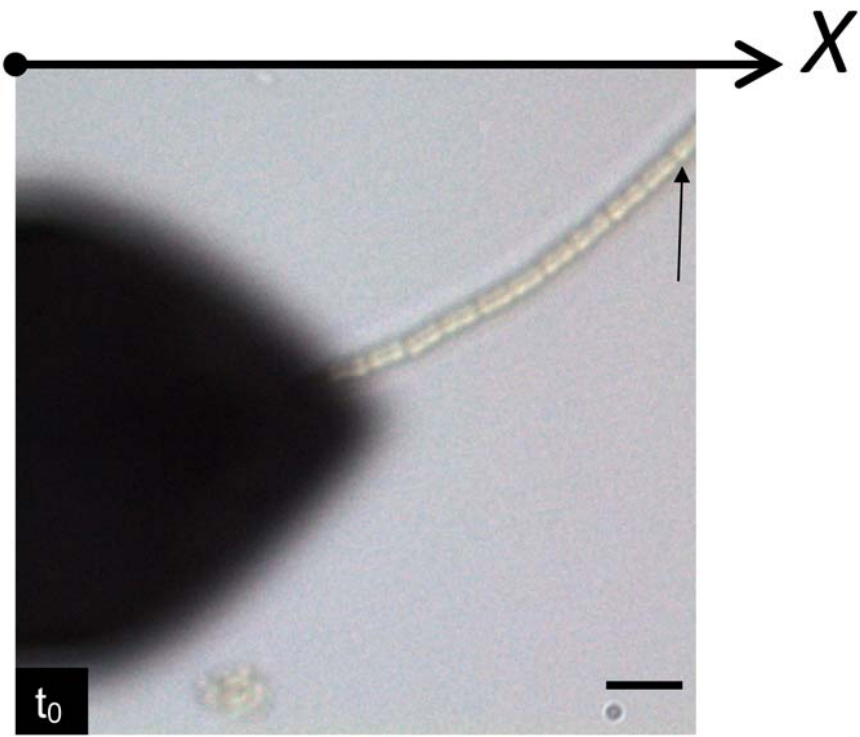

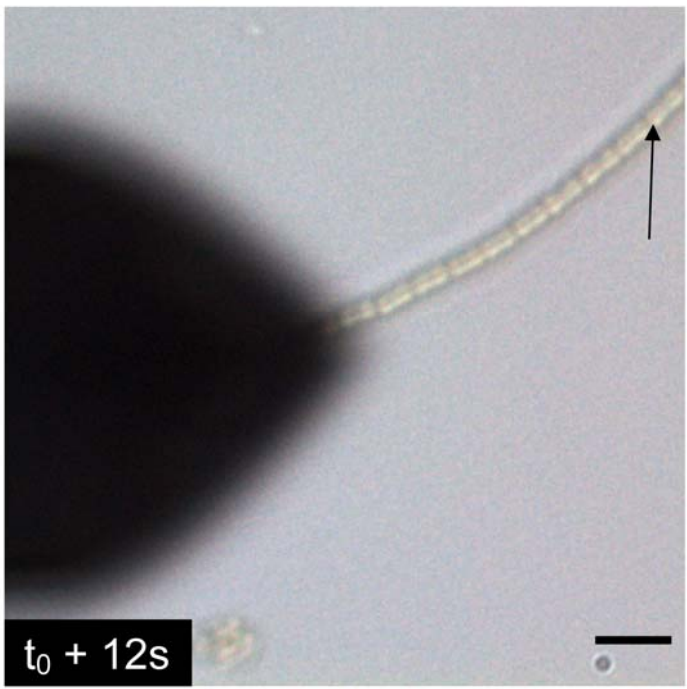

Figure 5 (version R2)

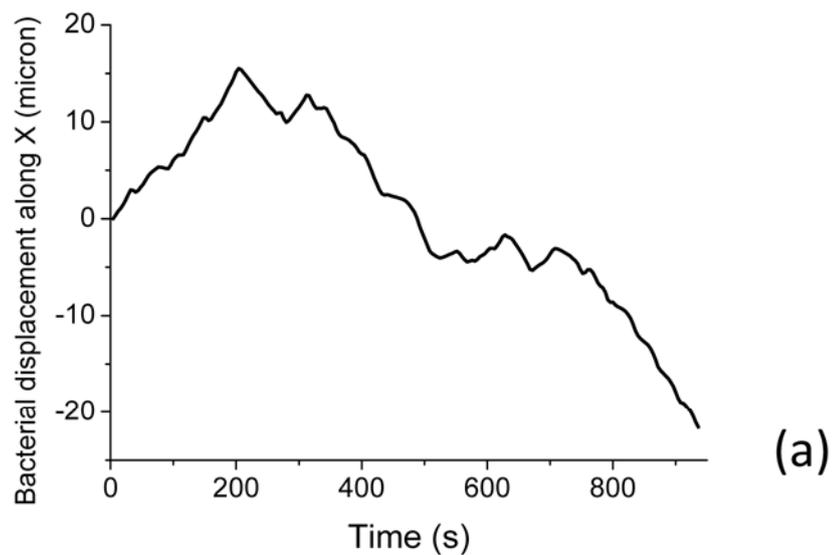

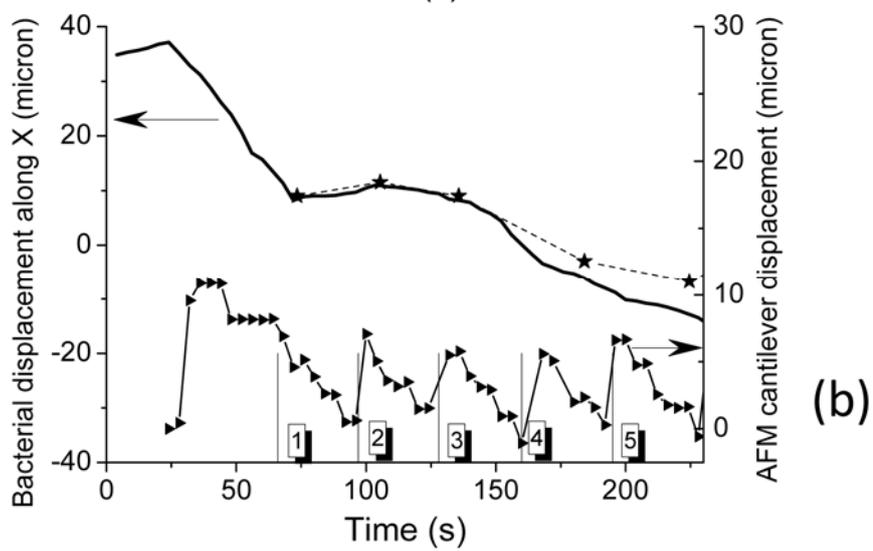

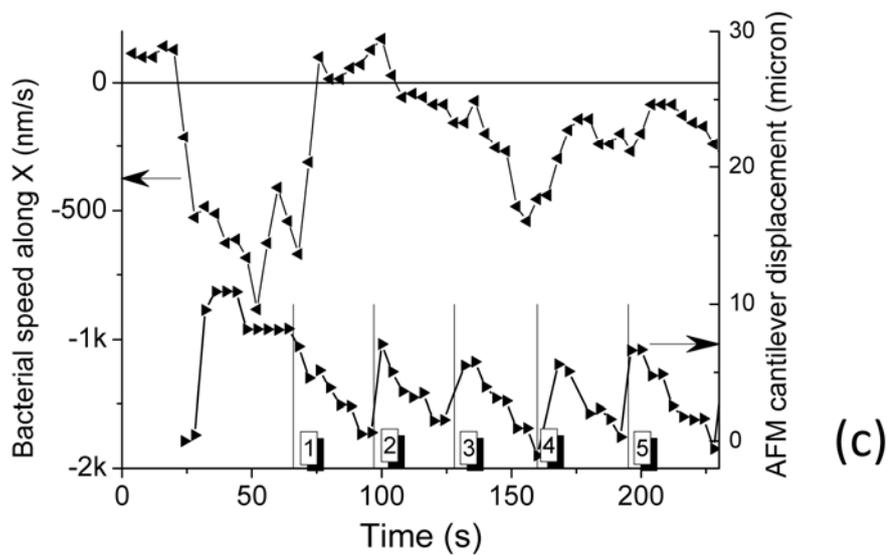

Figure 6 (version R2)

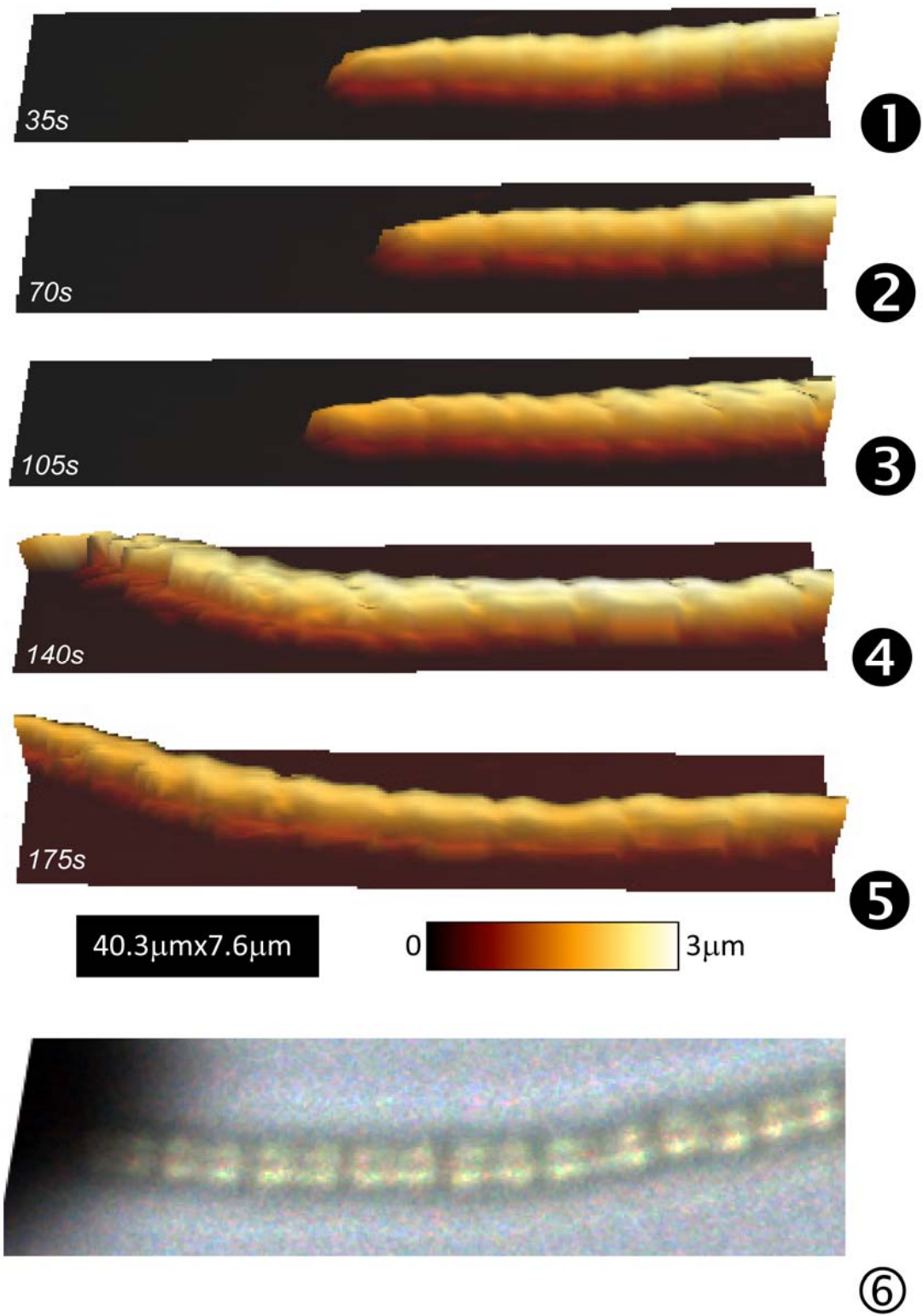

Figure 7 (version R2)

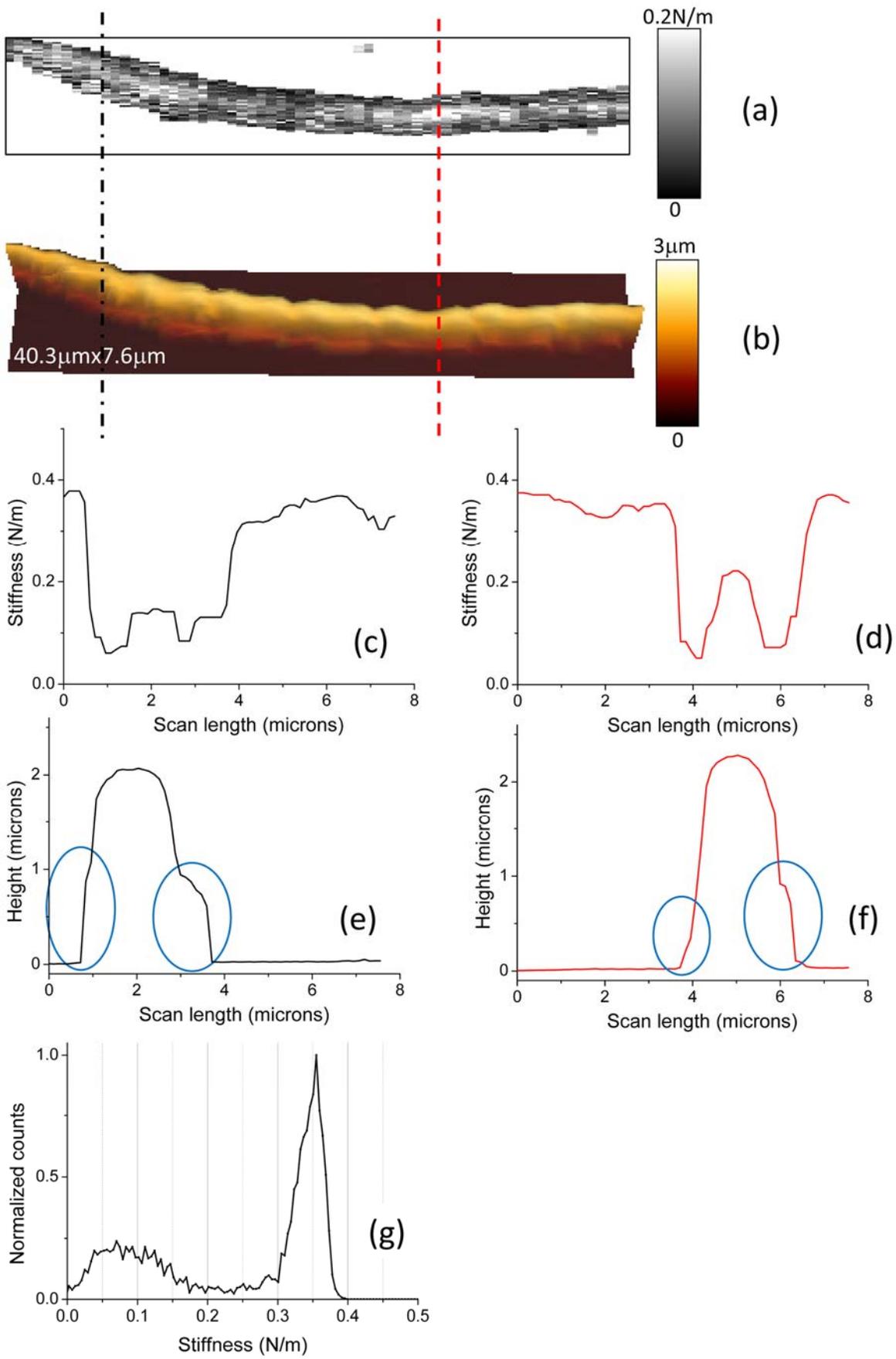

Figure 8 (version R2)

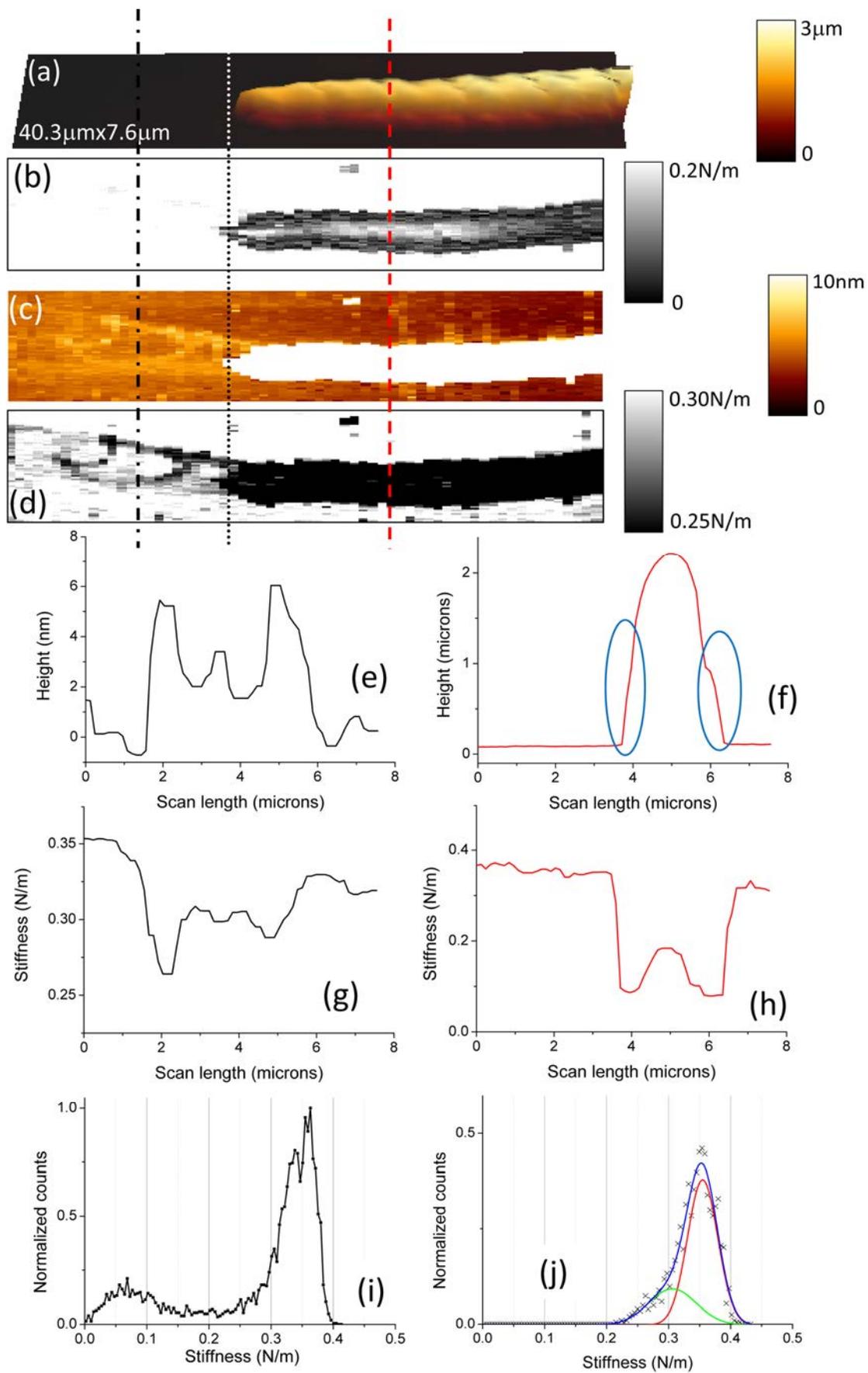

Figure 9 (version R2)

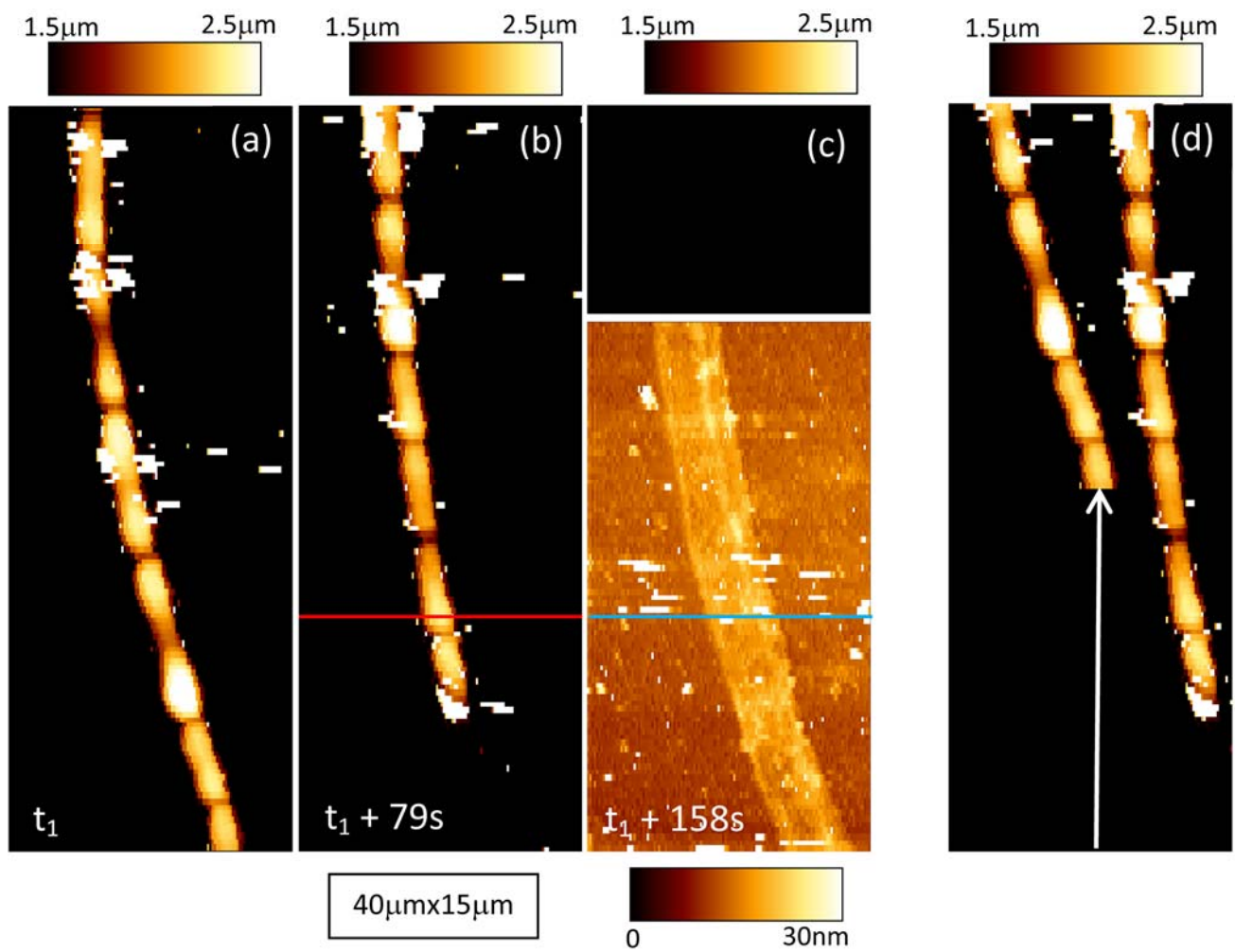

Figure 10 (version R2)

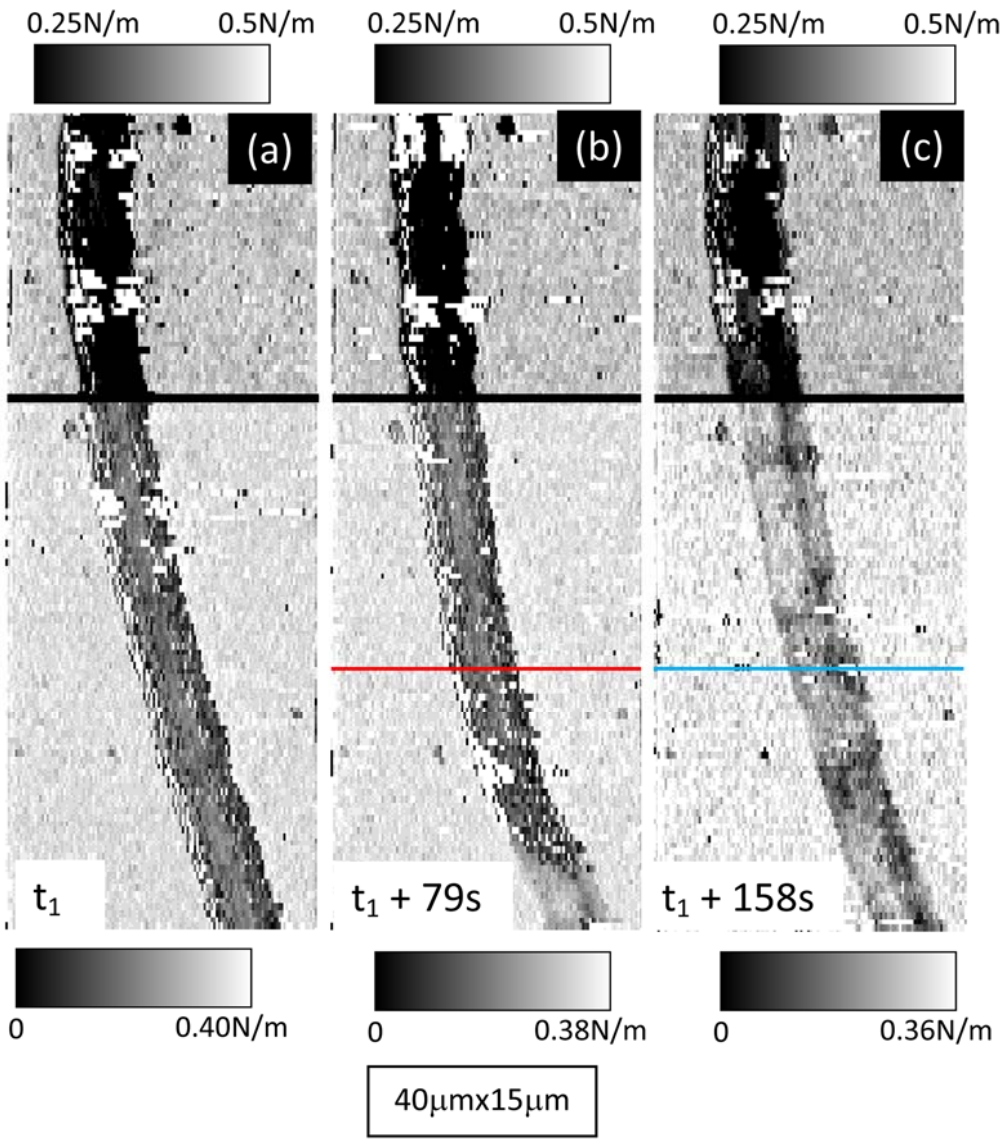
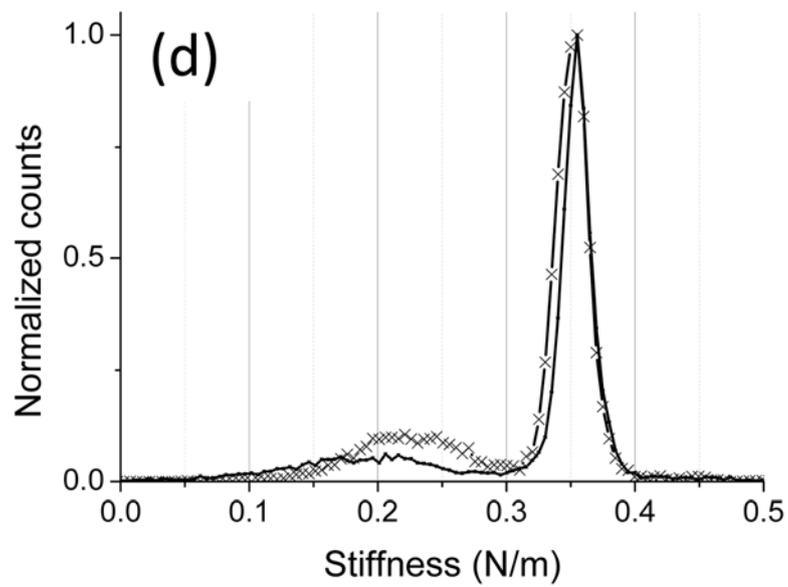

Figure 11 (version R2)

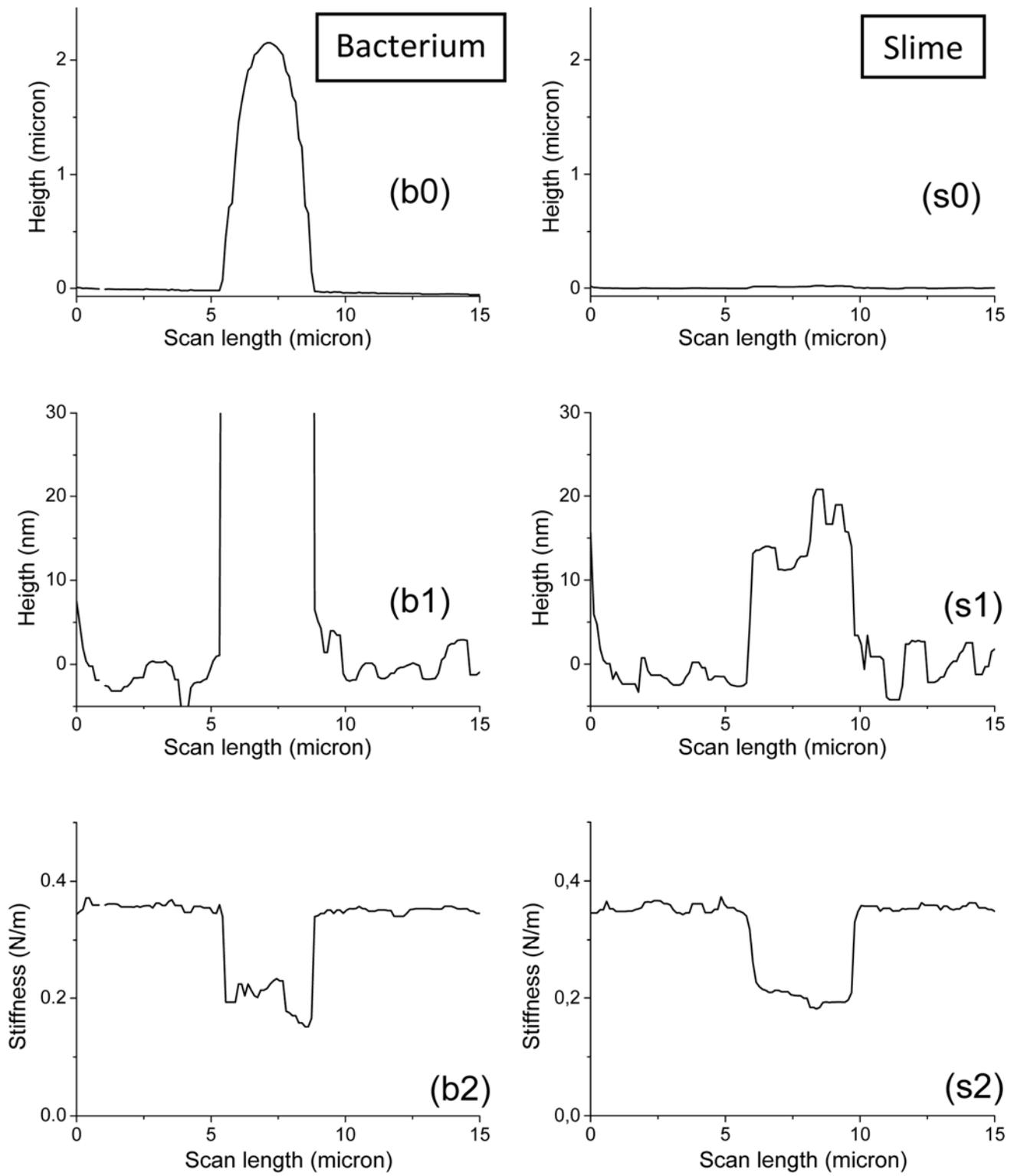

Figure 12 (version R2)

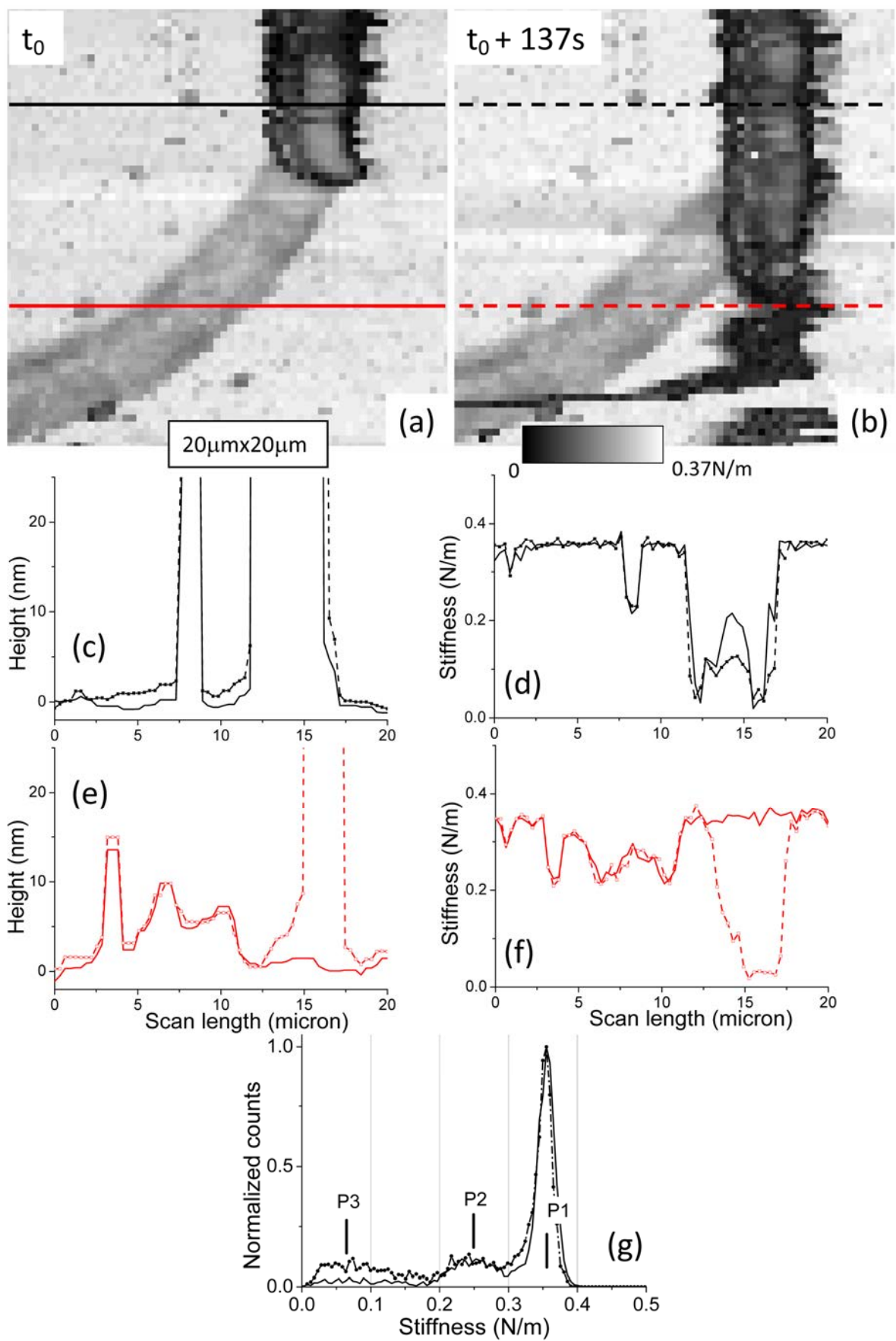

Figure 13 (version R2)

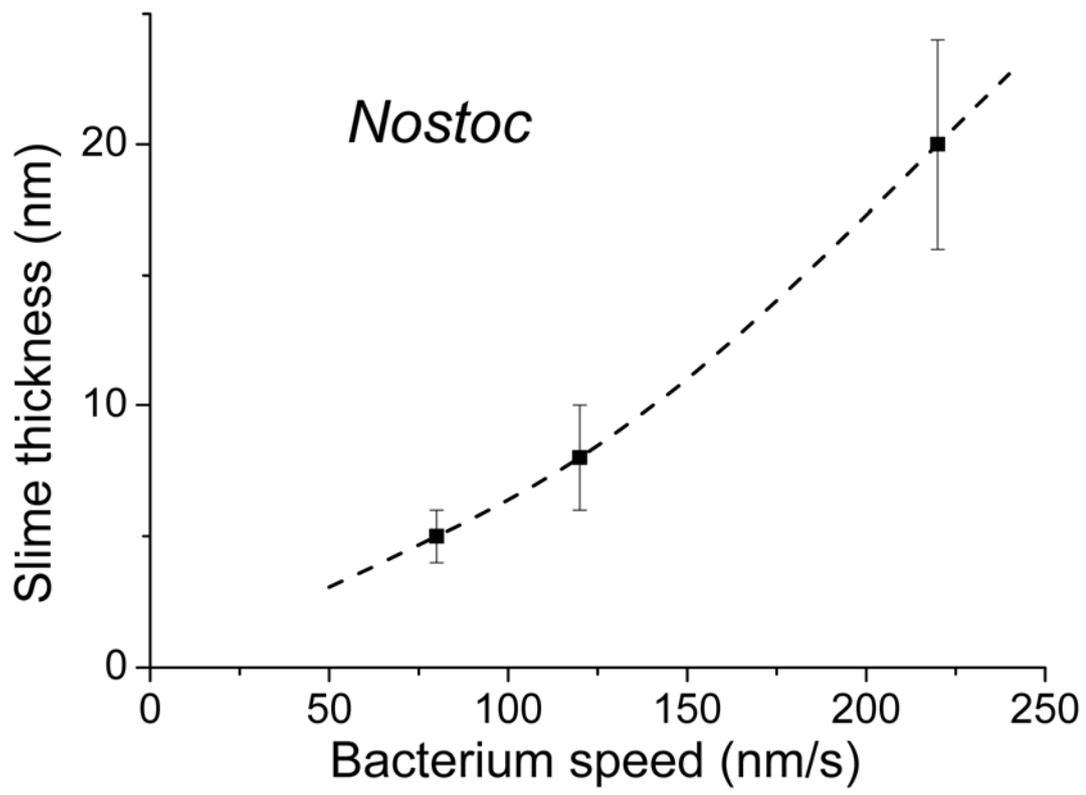

Figure 14 (version R2)

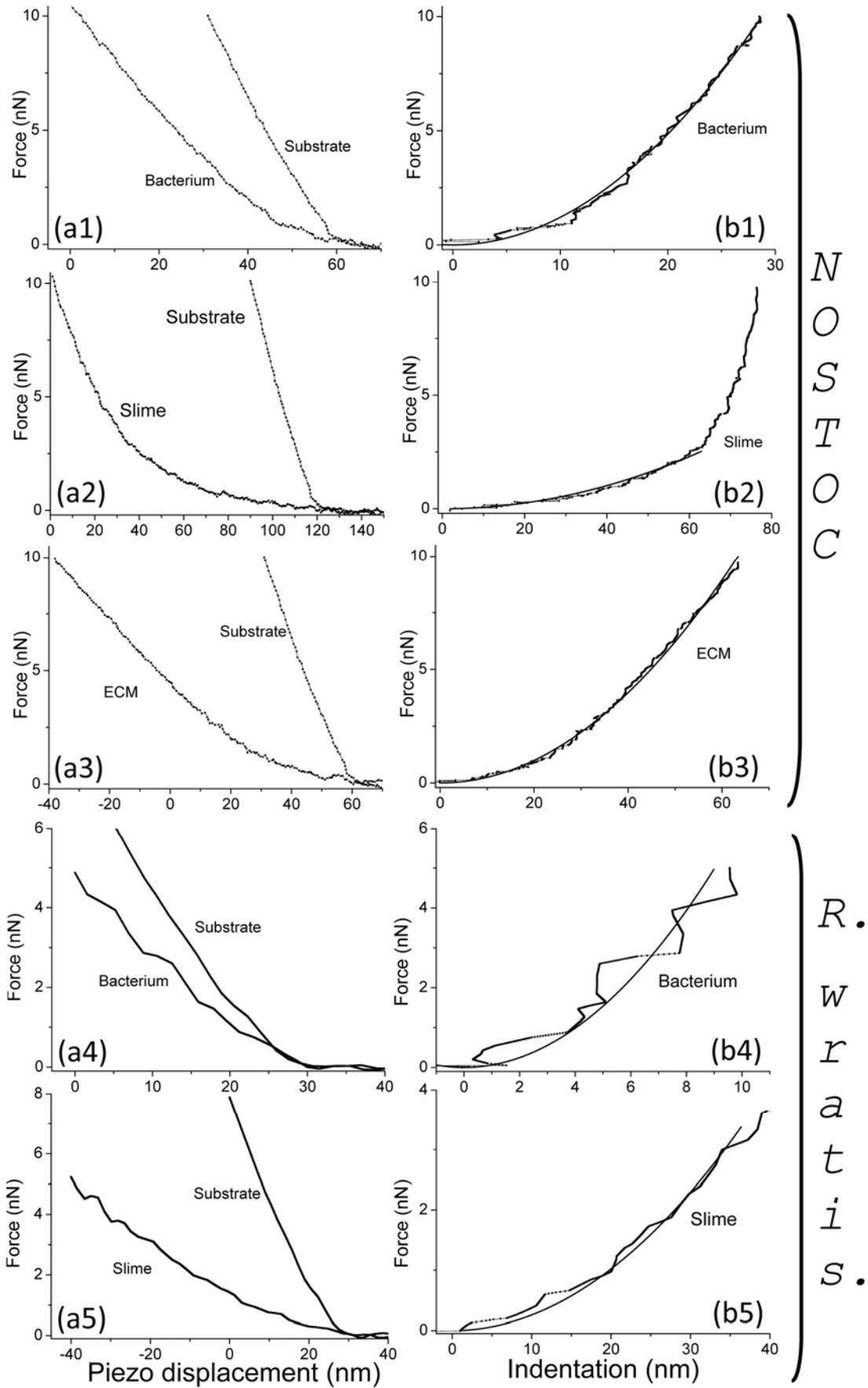

Figure 15 (version R2)